\documentclass[twocolumn, nofootinbib, 
prd, floats, floatfix, amsmath, amssymb, 
superscriptaddress, preprintnumbers]{revtex4-2} %
\UseRawInputEncoding
\usepackage[final]{graphicx}
\usepackage{hyperref}
\usepackage{amsmath}
\usepackage{bbm}
\usepackage{bm}
\usepackage{amsfonts}
\usepackage{amssymb}
\usepackage{latexsym}
\usepackage{graphicx}
\usepackage[english]{babel}
\usepackage{multirow}
\usepackage{float}
\usepackage{url}
\usepackage{slashed}
\usepackage{xcolor} 
\usepackage[utf8]{inputenc}
\usepackage{stmaryrd} 
\usepackage{enumitem}
\usepackage{hyperref}
\usepackage{cleveref}
\usepackage{siunitx}
\usepackage{verbatim}
\usepackage{orcidlink}
\usepackage{amsthm}

\theoremstyle{remark}

\newcommand{\be}{\begin{equation}}
\newcommand{\ee}{\end{equation}}
\newcommand{\ba}{\begin{array}}
\newcommand{\ea}{\end{array}}
\newcommand{\bea}{\begin{eqnarray}}
\newcommand{\eea}{\end{eqnarray}}

\newcommand{\besub}{\begin{subequations}}
\newcommand{\eesub}{\end{subequations}}

\newcommand{\nnn}{\nonumber \\}

\newcommand{\eg}{e.g.,\ }	
\newcommand{\ie}{i.e.,\ }	

\newcommand{\km}{{\rm km}} 
\newcommand{\s}{{\rm s}} 

\newcommand{\DM}{{\rm DM}} 
\newcommand{\beq}{\begin{equation} \begin{aligned}}
		\newcommand{\eeq}{\end{aligned} \end{equation}}

\definecolor{darkerblue}{rgb}{0.2,0.2,0.5}
\definecolor{seagreen}{rgb}{0.180392,0.545098,0.341176}
\definecolor{smagenta}{rgb}{0.5,0.145098,0.341176}
\definecolor{deepblue}{rgb}{0,0,1}

\begin{document}

\title{In situ Measurements of Dark Photon Dark Matter Using Parker Solar Probe: \\  Going beyond the Radio Window}
	
	\author{Haipeng An}
	\email{anhp@mail.tsinghua.edu.cn}
	\affiliation{Department of Physics, Tsinghua University, Beijing 100084, China}
	\affiliation{Center for High Energy Physics, Tsinghua University, Beijing 100084, China}
	\affiliation{Center for High Energy Physics, Peking University, Beijing 100871, China}	
	
	\author{Shuailiang Ge \orcidlink{0000-0003-1063-2282}}
	\email{sge@pku.edu.cn}
	\affiliation{Center for High Energy Physics, Peking University, Beijing 100871, China}
	\affiliation{School of Physics and State Key Laboratory of Nuclear Physics and Technology, Peking University, Beijing 100871, China}
    \affiliation{Department of Physics, Korea Advanced Institute of Science and Technology (KAIST), Daejeon 34141, South Korea}

	\author{Jia Liu \orcidlink{0000-0001-7386-0253}}
	\email{jialiu@pku.edu.cn}
	\affiliation{School of Physics and State Key Laboratory of Nuclear Physics and Technology, Peking University, Beijing 100871, China}
	\affiliation{Center for High Energy Physics, Peking University, Beijing 100871, China}
	
	\author{Mingzhe Liu}
	\email{mingzhe.liu@obspm.fr}
        \affiliation{LESIA, Observatoire de Paris, Université PSL, CNRS, Sorbonne Université, Université de Paris, 5 place Jules Janssen, 92195 Meudon, France}
	\affiliation{Space Sciences Laboratory, University of California, Berkeley, CA 94720-7450, USA}

\begin{abstract}
Dark photon dark matter (DPDM) emerges as a compelling candidate for ultralight bosonic dark matter, detectable through resonant conversion into photons within a plasma environment. This study employs in-situ measurements from the Parker Solar Probe (PSP), the first spacecraft to venture into the solar corona, to probe for DPDM signatures.
The PSP in-situ measurements go beyond the traditional radio window, spanning frequencies between about 10 kHz and 20 MHz, a challenging range inaccessible to Earth-based radio astronomy. Additionally, the proximity of PSP to the resonant conversion location enhances the signal flux, providing a distinct advantage over ground-based observations.
As a result, the PSP data establishes the most stringent constraints on the kinetic mixing parameter $\epsilon$ for DPDM frequencies between 70 kHz and 20 MHz, with values of $\epsilon \lesssim 10^{-14}-10^{-13}$. Investigating the data from STEREO satellites resulted in weaker constraints compared to those obtained from PSP. By utilizing state-of-the-art solar observations from space, we have surpassed the cosmic microwave background limits 
derived from early-universe observations.
\end{abstract}
\maketitle

\section*{Introduction}

\noindent
Dark matter, constituting approximately a quarter of the total energy in the present-day Universe, remains an enigma shrouded in mystery. The dark photon, also known as the hidden photon, emerges as a compelling dark matter candidate. It is a massive vector boson associated with an additional $U(1)$ gauge group~\cite{Holdom:1985ag, Dienes:1996zr, Abel:2003ue, Abel:2006qt}, often considered one of the simplest extensions to the Standard Model of particle physics. The production of the appropriate abundance of dark photon dark matter (DPDM) in the early Universe can occur through various mechanisms, including the enhanced misalignment mechanism\cite{Nelson:2011sf, Arias:2012az, AlonsoAlvarez:2019cgw, Nakayama:2019rhg, Nakayama:2020rka}, inflationary fluctuations \cite{Graham:2015rva, Ema:2019yrd, Kolb:2020fwh, Salehian:2020asa, Ahmed:2020fhc, Nakai:2020cfw, Nakayama:2020ikz, Kolb:2020fwh, Salehian:2020asa, Firouzjahi:2020whk, Bastero-Gil:2021wsf, Firouzjahi:2021lov, Sato:2022jya}, parametric resonances \cite{Co:2018lka, Dror:2018pdh, Bastero-Gil:2018uel, Agrawal:2018vin, Co:2021rhi, Nakayama:2021avl, Cyncynates:2023zwj} and the decay of cosmic strings \cite{Long:2019lwl}.

Dark photons, similar to ordinary photons, interact only weakly with the Standard Model. They can exhibit a natural kinetic mixing with photons, mediated by a small coupling constant~\cite{Holdom:1985ag, Dienes:1996zr, Abel:2003ue, Abel:2006qt}. This mixing allows for converting dark photons into photons and vice versa. This conversion process can occur in plasmas, where photons acquire an effective mass through collective plasma oscillations. When the dark photon mass coincides with the effective mass of a photon in a plasma, the conversion probability reaches a maximum, a phenomenon known as resonant conversion. With its vast expanse of plasma, the Sun provides an ideal environment for this process. As dark photons traverse the solar plasma, they undergo resonant conversion into photon flux with a certain probability, particularly when the dark photon mass aligns with the local plasma frequency at specific locations. The resulting photon signal exhibits a nearly monochromatic energy equal to the dark photon mass. Additionally, the signal flux benefits from the Sun's proximity compared to other distant astrophysical objects.

Previous studies~\cite{An:2020jmf, An:2023wij} have proposed employing terrestrial radio telescopes to search for such converted photon signals originating from the solar corona, leading to stringent constraints on the dark photon kinetic mixing coupling.
However, as ground-based radio telescopes, their observable frequency range is constrained by the radio window. For instance, radio frequencies below approximately 10 MHz are reflected back into space by the Earth's ionosphere, rendering these telescopes unable to provide any constraints. There exist well-motivated models, such as those presented in \cite{AlonsoAlvarez:2019cgw, Nakayama:2019rhg, Kolb:2020fwh, Nakai:2020cfw, Dror:2018pdh, Bastero-Gil:2018uel, Agrawal:2018vin, Cyncynates:2023zwj}, which predict DPDM around MHz scale.

To go beyond the radio window, we have employed a novel \textit{in-situ} search for DPDM, utilizing the Parker Solar Probe (PSP)~\cite{2016Fox} to directly measure the converted monochromatic photons in the solar plasma environment. Solar plasma extends well beyond the solar corona, reaching Earth's orbit at 1 AU and even further as the solar wind. Therefore, a solar probe immersed in the solar plasma can provide direct (\textit{in-situ}) measurements of the resonant conversion of dark photons into photons. The PSP stands as a remarkable achievement, being the first and only human-made spacecraft to venture into the solar corona~\cite{PSPorbitInfo}. It orbits around the Sun in highly elliptical trajectories, starting from its launch position on the Earth and reaching the closest perihelion distance of $\sim 10R_{\odot}$ from the Sun. The highly elliptical orbits allow PSP to probe a wide range of dark photon mass values that correspond to the plasma frequencies between the Sun and Earth. 

In addition to avoiding the constraints of the radio window, \textit{in-situ} measurements within the solar plasma offer yet another key advantage. Due to PSP's close proximity to the signal source, the photon flux converted from DPDM is significantly amplified compared to terrestrial radio telescopes. This is because the converted flux does not suffer from the attenuation caused by the vast distance between the Sun and Earth. 

Besides PSP, we have also utilized data from the Solar Terrestrial Relations Observatory (STEREO)~\cite{2008Kaiser_SSRv} to constrain the kinetic mixing coupling. STEREO consists of two spacecraft, one orbiting ahead of Earth (STEREO-A) and the other trailing behind (STEREO-B). However, the constraints derived from STEREO are not as stringent as those from PSP, primarily due to their fixed orbital distance of 1 AU from the Sun. This is true except for the low-frequency region below 70 kHz, where STEREO data can provide better constraints because such frequency aligns with the plasma frequency at its orbits. However, Cosmic Microwave Background (CMB) observations~\cite{Arias:2012az, Witte:2020rvb} already offer better exclusion than STEREO.

Consequently, the in-situ measurements of PSP effectively transform the solar corona into a dark matter haloscope. By utilizing the PSP observations, we present the search for DPDM in the radio frequency range from 70 kHz to 20 MHz, establishing the most stringent upper limits for the dark photon kinetic mixing coupling, surpassing the cosmological constraints from CMB observations.

\section*{Resonant conversion into photons}

The dark photon kinetically mixes with the ordinary photon through the term $\epsilon F^{\mu\nu} F'_{\mu\nu}$, where $F$ and $F'$ represent the field strengths of the photon and dark photon, respectively, and $\epsilon$ is the kinetic mixing parameter. In plasmas, photons acquire a non-zero mass equal to the plasma frequency $\omega_p$,
\beq\label{eq:omega_p}
\omega_p =\left( \frac{4\pi\alpha_{\rm EM} n_e}{m_e} \right)^{\frac{1}{2}}
\approx
10^{-8}~{\rm eV} \left( \frac{n_e}{7.3\times 10^4~{\rm cm}^{-3}}\right)^{\frac{1}{2}} \ ,
\eeq
where $\alpha_{\rm EM}$ denotes the fine-structure constant, $n_e$ represents the number density of electrons within the plasma, and $m_e$ signifies the electron mass. 
Throughout this work, we adopt the natural unit system for convenience where $\varepsilon_0 = c = \hbar = 1$ ($\varepsilon_0$ is the vacuum permittivity, $c$ is the speed of light, and $\hbar$ is the reduced Planck constant).
The probability for a dark photon $A'$ resonantly converting into a photon $\gamma$ is~\cite{An:2020jmf, Raffelt:1987im}
\beq\label{eq:prob}
P_{A'\rightarrow \gamma}
\simeq
\frac{2}{3} \pi \epsilon^2 m_{A'} \frac{1}{v_r(r_c)} \left|\frac{d\ln\omega_p^2(r)}{dr}\right|^{-1}_{r=r_c}.
\eeq
The resonant conversion occurs when the dark photon mass $m_{A'}$ equals the plasma frequency $\omega_{p}(r_c)$ at a specific radius $r_c$. In the context of this work, $\omega_p(r)$ represents the profile for the solar plasma frequency, extending beyond $1~{\rm AU}$ from the Sun. In Eq.~\eqref{eq:prob}, $v_r$ denotes the radial velocity of the dark photon. As demonstrated in Eq.~\eqref{eq:prob}, the conversion probability solely depends on the radial profile $\omega_p(r)$. Next, we show more details of deriving the flux of converted photons in the context of the solar plasma considered in the present work. 

The DPDM can reach the resonant layer $r_c$ if its impact parameter $b$ falls within the maximum value, $b_{\rm max}= r_c v(r_c)/v_0$. By integrating over all incident DPDM from $b=0$ to $b_{\rm max}$, we obtain the total power of converted photons emanating from a sphere with radius  $r_c$~\cite{An:2020jmf}:
\beq \label{eq:tot_power}
\mathcal{P}_0 = 4\pi r_c^2  P_{A'\rightarrow \gamma}(v_0) \rho_{\DM} v(r_c) 
\eeq
where $P_{A'\rightarrow \gamma}(v_0)$ is the conversion probability defined in Eq.~\eqref{eq:prob}, replacing $v_r(r_c)$ with the initial velocity of the dark photon, $v_0$. Both incoming and outgoing DPDMs passing through the resonant layer contribute to the flux of converted photons, resulting in a doubling of the power. This occurs because photons converted from the incoming DPDM will be totally reflected due to the higher plasma frequency within the inner sphere. This effect has already been accounted for in Eq.~\eqref{eq:tot_power}. In realistic calculations, we also consider the distribution of the initial DPDM velocity. Consequently, the total converted power has to be averaged over the DPDM velocity profile as $\mathcal{P}=\int_{0}^{\infty} dv_0 ~\mathcal{P}_0(v_0) f_{\rm DM}(v_0)$. A detailed description of this averaging process is provided in the Supplemental Material~\cite{Supp_cite}.
~\nocite{Witte:2021arp, Mirizzi:2009iz, Drukier:1986tm,Choi:2013eda,Evans:2018bqy, Hardy:2022ufh, favorite2016solid, manning2000instrumentation, zarka2004jupiter, eastwood2009measurements, PSPdataLevel3, Moncuquet2020First/FIELDS, balanis2015antenna, Maksimovic2020AnticorrelationHelios, SQTN_WEB, PSP_FIELDS_Release_Note, 2017meyervernet, Kasper2016, Maksimovic2020,2022Martinovic,2023Liu_QTN,Zheng_2024, Case2020, 2022Livi, Whittlesey_2020, 2021Kasper,2021Zhao,2021Liu,2021ApJLiu,2024LiuYingD, meyer1989tool, An:2024kls, 
1989ApJ...337.1023C, 2001SSRv...97....9W, 2007ApJ...671..894T, 2008ApJ...676.1338T, 2018ApJ...857...82K,
2020ApJS..246...57K,
Brahma:2023zcw}

Then, the spectral flux density at the location of a spacecraft or satellite situated at a distance $R$ from the Sun is given by
\beq \label{eq:S_sig}
S_{\rm sig} = \frac{1}{2} \frac{1}{4\pi R^2} \frac{1}{\mathcal{B}}\mathcal{P}.
\eeq
The prefactor $1/2$ is due to averaging over the polarizations of propagating photons.
If $R$ is close to $r_c$, the satellite can measure the resonant conversion of dark photons into photons directly (\textit{in-situ}), thereby eliminating the signal attenuation caused by the distance. The location range of the PSP spans from approximately 1 AU to as close as $\sim 10 R_{\odot}$ from the solar center. The electron density of the solar plasma from the solar corona to 1 AU can be well approximated by the following relation~\cite{1998SoPh..183..165L}:
\beq \label{eq:ne_profile}
& n_e(r) =  \frac{n_e(1{\rm~AU})}{7.2}
\left[3.3\times10^5 \left(\frac{r}{R_{\odot}} \right)^{-2} \right.
\\
& 
\left.
+ 4.1\times 10^6 \left(\frac{r}{R_{\odot}} \right)^{-4} 
+ 8.0\times 10^7 \left(\frac{r}{R_{\odot}} \right)^{-6} \right] \ .
\eeq
The electron density at $1~{\rm AU}$, $n_e(1{\rm~AU})$, is approximately $10~{\rm cm}^{-3}$.  
Therefore, according to the orbits of PSP and Eq.~\eqref{eq:omega_p}, PSP will traverse the plasma mass ranging from 
$10^{-10}$ eV to $ 3\times 10^{-9}$~eV
(or equivalently, the plasma frequency from $150$~kHz to $4500$~kHz).
For heavier dark photon mass (corresponding to smaller resonant radius $r_c$), the converted normal photons can still be detected by PSP, although weakened by distance attenuation. However, for resonant conversion layers farther than the satellite's farthest position, the signal cannot reach the satellite due to the inner plasma shielding effect, because the inner plasma has a higher plasma frequency than the signal frequency.
In addition, the small-scale density fluctuations in the solar plasma do not alter our results significantly, which is demonstrated in detail in the Supplemental Material~\cite{Supp_cite}.

\section*{PSP constraints}

Due to its highly elliptical orbit, PSP can provide \textit{in-situ} measurements to the photons from resonant conversion of DPDM, within the solar plasma between the Sun and the Earth. The FIELDS instrument, one of the primary instruments aboard the PSP, houses a Radio Frequency Spectrometer (RFS) equipped with two radio receivers: the low-frequency receiver (LFR) monitors emissions in the frequency range of 10 kHz to 1.7 MHz, while the higher-frequency receiver (HFR) captures emissions in the range of 1.3 MHz to 19.2 MHz~\cite{Pulupa2017TheProcessing}.

Since its launch on August 12, 2018, PSP has completed over ten orbits around the Sun. We will utilize the publicly available data collected until July 2022 by the two radio receivers, LFR and HFR.
Since PSP has approached nearly the closest distance $\sim 10~R_{\odot}$ by July 2022, including more data collected after this time will not further significantly enhance our results. 

For the sake of efficient data analysis, we define an ``orbital phase" as a full orbit spanning two consecutive aphelion dates marking its initiation and termination. The orbital phase information is summarized in Table~I in the Supplemental Material~\cite{Supp_cite}. In total, we possess 1151 LFR data files and 1154 HFR data files collected on the dates spanning from October 02, 2018, to July 19, 2022.
Initially, we must calibrate the recorded data to convert its units to those of spectral flux density, consistent with the unit of the signal~\eqref{eq:S_sig}. We then filter the data by removing significant time-dependent noises. This will enhance data quality while simultaneously preserving the constant DPDM-induced signal. The information regarding PSP positions on various days is available here~\cite{PSPwebPosition}. Moreover, PSP provides \textit{in-situ} measurements of $n_e$, and the archived data can be accessed here~\cite{2020Moncuquet, PSP_QTN_CNES}. Since the recorded data for each day is used to provide constraints across a broad range of frequencies, the measured data may not be sufficient to determine the local $n_e$ information at the resonant radius for all these frequencies. Therefore, we still rely on Eq.~\eqref{eq:ne_profile} and use the measured data to adjust the normalization of $n_e$ at 1 AU, as detailed in the Supplemental Material~\cite{Supp_cite}.

Regarding the data analysis and the calculation of constraints on the kinetic mixing, we generally follow the procedure outlined in our previous work~\cite{An:2022hhb, An:2023wij}. For each frequency bin in the observed data, we calculate the averaged spectral flux density $\bar{O}_i$ with the uncertainty
$\sigma_{\bar{O}_i}$. To effectively utilize these data in constraining the kinetic mixing parameter $\epsilon$, we employ a likelihood-based statistical method~\cite{Cowan2011}. First, we fit the data background locally around the frequency bin $i$ and its neighboring bins using a polynomial function. We also adopt the method in~Ref.~\cite{ParticleDataGroup:2018ovx} to rescale the errors, in order to avoid underestimation of the errors.
Then, we construct a likelihood function $L$ that relates the fitting function to the data background with the DPDM-induced signal included, assuming its oscillation frequency falls within the frequency bin $i$. Since DM is non-relativistic and has a velocity of approximately $v_{\rm DM}\sim  10^{-3}c$, its energy spread is only about $10^{-6}$. Because the bin size ranges from 8 kHz to 200 kHz which is much larger than the signal width, the observed signal is well-contained within the $i$th frequency bin.

\begin{figure}
    \centering
    \includegraphics[width=1\linewidth]{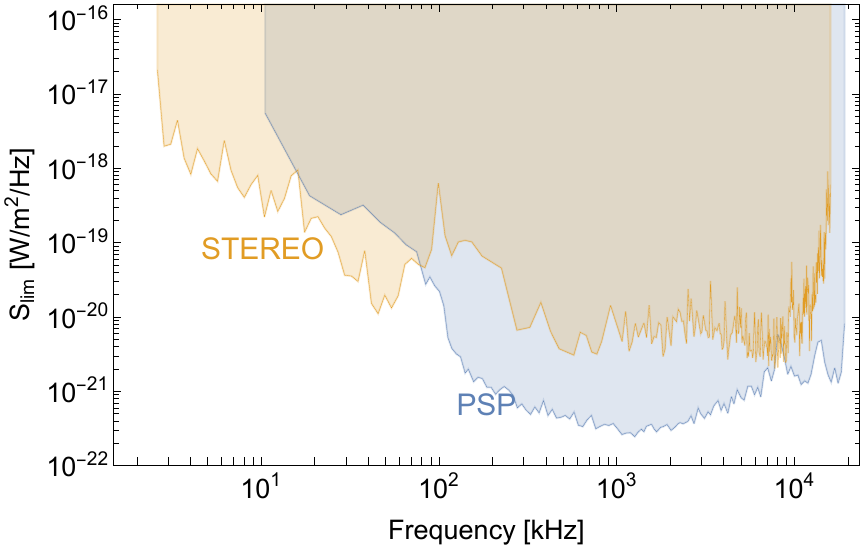}
    \caption{The 95$\%$ C.L. upper limits on the model independent monochromatic photon flux $S_{\rm lim}$. The shaped blue region represents the constraints from PSP data, plotted by selecting the strongest constraint for each frequency bin from all available days between October 02, 2018, and July 19, 2022.
    The shaped orange region shows the constraint from STEREO data on January 13, 2007.}
    \label{fig:S_lim}
\end{figure}

We define the global maximum $L_G$ of the likelihood function, achieved by simultaneously optimizing both the nuisance parameters in the fitting function and the signal amplitude $S_{\rm sig}$. Additionally, we define the conditional maximum $L_C$, obtained by optimizing only the nuisance parameters in the fitting function while keeping $S_{\rm sig}$ fixed. The test statistic $-2\ln(L_C/L_G)$, as a function of $S_{\rm sig}$, follows the half-chi-square distribution~\cite{Cowan2011}. Utilizing this relationship, we calculate the $95\%$ confidence level (C.L.) upper limits on $S_{\rm sig}$ for each data file. Detailed information about the data analysis process, including data calibration, data filtering, and the statistical method, is provided in the Supplemental Material~\cite{Supp_cite}.

\begin{figure}
    \centering
    \includegraphics[width=1\linewidth]{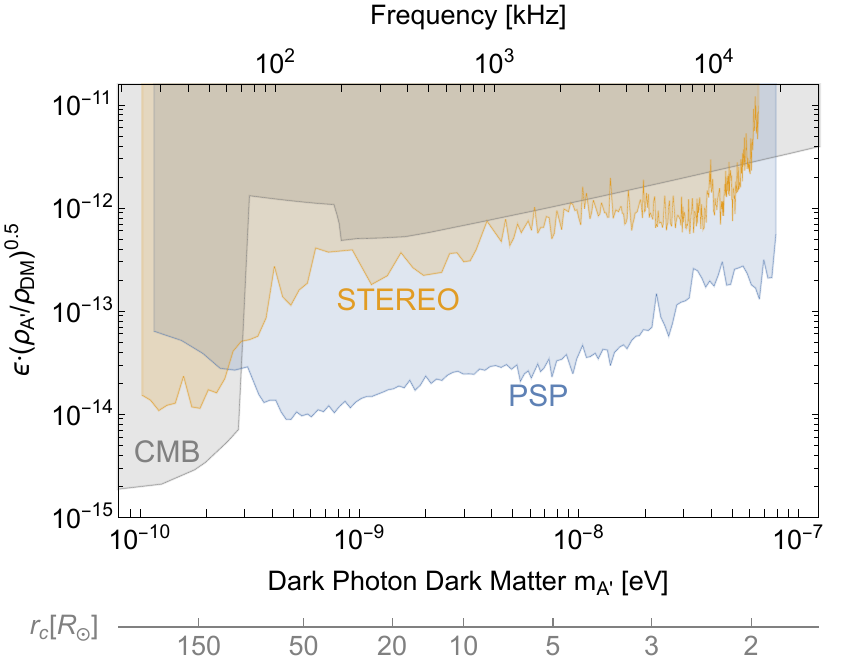}
    \caption{The 95$\%$ C.L. upper limits on the kinetic mixing parameter $\epsilon$, scaled with the square root of the fraction of the local DPDM density. The shaped blue region represents the constraint from PSP data, while the shaded orange region represents the constraint from STEREO data, similar to Fig.~\ref{fig:S_lim}. Additionally, we show the existing constraints from CMB observations~\cite{Arias:2012az, Witte:2020rvb} in the shaded gray region.
    A third horizontal axis, $ r_c $, is provided to indicate the location of resonant conversion, where $ \omega_p(r_c) = m_{A'} $.
    }
    \label{fig:constraint_eps}
\end{figure}

Finally, we establish the flux limits on the injection of radio photons into a single frequency bin using PSP data, denoted as \(S_{\text{lim}}\). These limits are independent of the details of the DM model, as long as the received photons are monochromatic.
Fig.~\ref{fig:S_lim} presents the envelope of $S_{\text{lim}}$ constraints derived from all 2305 data files collected on the days listed in Table~I in the Supplemental Material~\cite{Supp_cite}. To obtain this result, we have selected the strongest limit on $S_{\rm lim}$ for each frequency bin among all available days. The frequency range encompasses that of both receivers, LFR and HFR. Additionally, we include the constraint obtained from STEREO data, following the same data analysis procedures. Since the STEREO orbit remains nearly stationary at a distance of $1~{\rm AU}$ from the Sun, we only utilize data from one day, January 13, 2007. On this day, the Sun exhibits relatively quiet conditions, and the data has been utilized for data calibration~\cite{zaslavsky2011antenna}. We emphasize that Fig.~\ref{fig:S_lim} represents model-independent constraints on a constant monochromatic signal from both PSP and STEREO data. These constraints can be applied to restrict any new physics models that could potentially generate an excessive monochromatic radio signal that would be detectable by these probes.

Afterward, we convert the flux constraints on $S_{\rm lim}$ into constraints on the kinetic coupling $\epsilon$ using Eq.~\eqref{eq:S_sig}, and the results are presented in Fig.~\ref{fig:constraint_eps}. 
When creating Fig.~\ref{fig:constraint_eps}, we have excluded constraints at frequencies below the local plasma frequency (divided by $2\pi$) for each day. Signals at such low frequencies may be generated by DPDM at a distance from the Sun farther than the satellite position, but they cannot propagate inward to be detected by the satellite.
The figure depicts the envelope of the constraints, where we have selected the strongest limit for each frequency bin from all available days. The figure demonstrates the upper limit on $\epsilon$ derived using PSP data in blue shaded region, which surpasses the CMB constraint~\cite{Arias:2012az, Witte:2020rvb} in gray shaded region, by approximately 1 to 2 orders of magnitude in the frequency range $f$ between 70~kHz and 20~MHz or equivalently in the range of dark photon mass $m_{A'}$ between $3\times 10^{-10}$ and $8\times 10^{-8}$~eV, where $f=m_{A'}c^2/(2\pi \hbar)$.
We also include the constraints obtained from STEREO data on January 13, 2007.  PSP outperforms STEREO at higher frequencies due to its ability to directly observe (or at least be less affected by distance attenuation) the DPDM-induced signals in these frequencies as it travels closer to the Sun. This key advantage underscores the power of \textit{in-situ} measurements. Furthermore, the frequency ranges tested can extend beyond the radio window, offering a substantial advantage over terrestrial experiments.

\begin{figure}
    \centering
    \includegraphics[width=\linewidth]{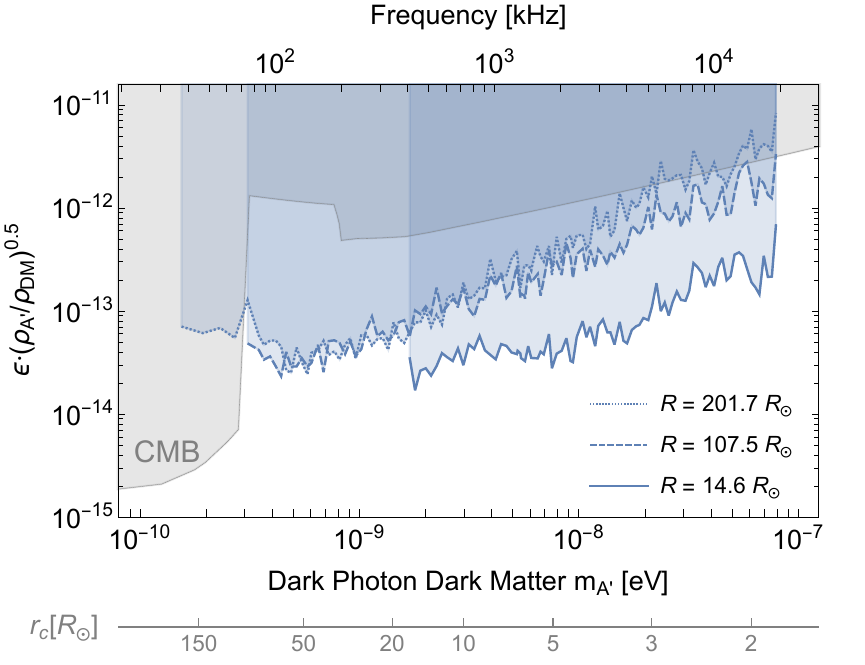}
    \caption{The 95$\%$ C.L. upper limits on the kinetic mixing parameter $\epsilon$ using the data collected by PSP at different distances $R$ from the solar center on various dates. The examples we choose are $201.7~R_{\odot}$ (November 15, 2019), $107.5~R_{\odot}$ (January 01, 2021) and $14.6~R_{\odot}$ (June 01, 2022), respectively, for illustrative purposes. 
    A third horizontal axis, $ r_c $, is provided for $ \omega_p(r_c) = m_{A'} $.
    }
    \label{fig:results_distances}
\end{figure}

Additionally, to better illustrate the advantages of the in-situ measurements by PSP, in Fig.~\ref{fig:results_distances}, we show the constraints on $\epsilon$ using data collected at different distances from the Sun on various dates. We see a clear benefit from the short distances between the satellite and the Sun, especially at high frequencies.
The constraints for each day exhibit a sharp cut-off on the left-hand side, positioned at the local plasma frequency (divided by $2\pi$). This cut-off arises because low-frequency signals cannot reach the PSP.

We can also translate the constraints on $\epsilon$ into the constraints on the coupling between axion dark matter and photons, $g_{a\gamma\gamma}$, via $\sqrt{2/3}\epsilon m_{A'}^2 \Leftrightarrow g_{a\gamma\gamma} |\boldsymbol{B}_T| m_a$ where $m_a$ is the axion mass and $\boldsymbol{B}_T$ is the solar magnetic field in the transverse direction~\cite{An:2023wij, An:2023mvf}. 
For example, for $m_{A'}\sim 10^{-7}$~eV, the resonant region is $r_c \sim 1.6 R_{\odot}$ corresponding to $|\boldsymbol{B}_T|\sim 0.3$~Gauss~\cite{2020Sci...369..694Y, An:2023wij}. Then, $\epsilon \sim 10^{-13}$  (see Fig.~\ref{fig:constraint_eps}) corresponds to $g_{a\gamma\gamma} \sim 10^{-9} {\rm GeV}^{-1}$ at $m_a\sim 10^{-7}$~eV, which cannot beat multiple existing constraints on axion~\cite{AxionLimits}. Such constraints on axion are expected to become even weaker for lower $m_a$ as the resonant conversion happens further away from the Sun where the magnetic field decreases quickly~\cite{Bale2016TheTransients}.

\section*{Conclusion}

We have established a novel \textit{in-situ} approach to probing the elusive DPDM by exploiting the resonant conversion of DPDM into monochromatic photons in the solar plasma environment, leveraging data from PSP. With the help of PSP, we have ventured into the unexplored realm of radio frequencies between 10 kHz and 20 MHz, previously inaccessible to ground-based radio observatories and other laboratory experiments. 
Due to the proximity to the resonant conversion location and the exceptional sensitivity of PSP, we have set the most stringent contemporary limits on the kinetic mixing parameter $\epsilon$ for radio-frequency dark photons with masses $m_{A'} \sim 3\times 10^{-10}-8\times 10^{-8}$~eV (70 kHz to 20 MHz). These limits, reaching down to $\epsilon \lesssim 10^{-14}-10^{-13}$, surpass even those established from early universe observations of the cosmic microwave background. The PSP mission, with its unique vantage point in the solar environment, provides an invaluable platform for exploring the properties of DPDM in the contemporary universe, complementing and enriching existing astrophysical and cosmological observations.

\section*{Acknowledgment}
We would like to thank Baptiste Cecconi, Arnaud Zaslavsky, and Marc Pulupa for helpful discussions on the STEREO and Parker Solar Probe data analysis. The work of HA is supported in part by the National Key R$\&$D Program of China under Grant No. 2023YFA1607104 and 2021YFC2203100, the National Science Foundation of China (NSFC) under Grant No. 12475107.  
The work of SG is supported by NSFC under Grant No. 12247147, the International Postdoctoral Exchange Fellowship Program, and the Boya Postdoctoral Fellowship of Peking University. 
The work of JL is supported by NSFC under Grant No. 12475103,  12235001 and 12075005.

\newpage

\setcounter{equation}{0}
\renewcommand{\theequation}{S\arabic{equation}}

\setcounter{figure}{0}
\renewcommand{\thefigure}{S\arabic{figure}}

\begin{center}
  {\large{\textbf{\textit{Supplemental Material}}}}  
\end{center}



\section{Conversion probability}

\noindent
For dark matter particles with typical initial velocity $v_0\simeq 220 ~ \km/\s$, the solar gravitational effect cannot be neglected when they travel close to the Sun since $GM_{\odot}/(R_{\odot} v_0^2)\simeq 3.45$ is an $\mathcal{O}(1)$ number. The trajectory of a dark matter particle is hyperbolic in the solar gravitational potential. We denote this trajectory as $l$, which can be expressed as 
\begin{align}
& \frac{dl}{dr}=\frac{1}{\cos\alpha(r)} = \pm \frac{v}{v_r},  
\label{eq:trajectory}
\\
& v_{\theta} = v_0 \frac{b}{r}
,~~~
v_{r} = \sqrt{v_0^2- v_{\theta} ^2 +\frac{2GM_{\odot}}{r}}
,~~~
v^2 = v_r^2 + v_{\theta}^2, \nonumber
\end{align}
where $r$ is the radial distance from the Sun, and $b$ is the impact parameter. $v_r$ and $v_{\theta}$ represent the radial and tangential velocities, respectively, and both are defined as positive quantities. The sign associated with the term $v/v_r$ in the first line, depends on the value of $\cos\alpha$, where $\alpha$ represents the angle between the velocity vector $\boldsymbol{v}$ and the radial vector $\boldsymbol{r}$. 

The Lagrangian of photon and dark photon in the original form with the kinetic mixing term can be written as
\bea
\mathcal{L} &=& -\frac{1}{4}F_{\mu\nu}F^{\mu\nu}-\frac{1}{4}F'_{\mu\nu}F'^{\mu\nu} + \frac{1}{2}m_{A'}^2 A'_{\mu}A'^{\mu} \nnn
&& - \frac{1}{2}\epsilon F'_{\mu\nu}F^{\mu\nu} + e A_\mu J^\mu,
\eea
where $e$ is the electromagnetic coupling and $J^\mu$ is the electric current. In the vacuum, since gravity couples to the mass, after universe-age-long evolution, the DPDM is in the mass eigenstate. Thus, we redefine the photon field as, 
\bea
A\rightarrow A - \epsilon A' \ .
\eea
Then the Lagrangian can be written as
\bea\label{eq:lag0}
\mathcal{L} &=& -\frac{1}{4}F_{\mu\nu}F^{\mu\nu}-\frac{1}{4}F'_{\mu\nu}F'^{\mu\nu} + \frac{1}{2}m_{A'}^2 A'_{\mu}A'^{\mu} \nnn
&& + (e A - \epsilon e A')_\mu J^\mu,
\eea
One can see that in (\ref{eq:lag0}), $A$ and $A'$ are diagonalized with the price of introducing the interaction between $A'$ and the electric current. 

Inside the plasma, the forward scattering of photon and dark photon off the charged particles induces a correction to the Lagrangian,
\bea
\frac{1}{2} (A - \epsilon A')_\mu \Pi^{\mu\nu} A_\nu (A - \epsilon A')_\nu \ ,
\eea
where in Fourier space the polarization tensor $\Pi^{\mu\nu}$ can be written as
\bea
\Pi^{\mu\nu} = \Pi_T \sum_i \varepsilon_{Ti}^\mu \varepsilon_{Ti}^\nu + \Pi_L \varepsilon_L^\mu \varepsilon_L^\nu \ . 
\eea
$\Pi_T = \omega_p^2$ and $\Pi_L = \omega_p^2 k^2/{k^0}^2$\ , where $k$ is the four-momentum in the Fourier space. Thus, for transverse components of $A$ and $A'$, we can derive the equation of motion,
\bea\label{eq:diff1}
\left[ \frac{\partial^2}{\partial t^2} - \nabla^2 + \left( \begin{array}{cc} \omega_p^2 ~&~ -\epsilon \omega_p^2 \\ -\epsilon \omega_p^2 ~&~ m_{A'}^2 \\ \end{array} \right) \right] \left( \begin{array}{cc} A \\ A' \\ \end{array}\right) = 0 \ .
\eea
To arrive at the above equation, we neglect the spatial variation of the plasma frequency. 

The variation can be quantified as 
\bea
\left|\frac{1}{\omega_p} \frac{d\omega_p}{d l}\right|_{r = r_c} = \left|\frac{1}{\omega_p} \frac{d\omega_p}{d r}\right|_{r = r_c} \approx \frac{1}{r_c} \ ,
\eea
where we have omitted ${\cal O}(1)$ factors and $r_c$ denotes the resonance radius. Based on the solar plasma profile given by Eq.~\eqref{eq:ne_profile}, we can demonstrate that in the region extending from $10 R_{\odot}$ to 1 AU, the following inequality holds:
\bea\label{eq:WKB1}
|k_c r_c| \gg 1,
\eea
where $k_c \equiv v_{\rm DM} \omega_p$ represents the momentum of dark photon dark matter particles (DPDM) that can resonantly convert into photons at the resonance radius $r_c$. Thus, it is legitimate to use the differential equation~\eqref{eq:diff1} to describe the evolution of the photon and dark photon fields. 

Eq.~\eqref{eq:WKB1} also allows to use the WKB method to solve \eqref{eq:diff1}. Following Ref.~\cite{Witte:2021arp}, we define 
\bea\label{eq:S10new}
A(t,l) &=& \frac{\tilde{A}(l)}{[k_A(l)]^{1/2}} \exp\left(- i\omega t + i \int^l_{l_0} k_A(l') dl'\right) \ , \nnn
A'(t, l) &=& \tilde{A}'_0 \exp(- i\omega t + i k l) \ ,
\eea
where $k = (\omega^2 - m_{A'}^2)^{1/2}$ is the momentum of $A'$ and $k_A(l) = (\omega^2 - \omega_p^2(l))^{1/2}$ is the position-dependent momentum of $A$. 
Then, substitute \eqref{eq:S10new} into the equation of motion \eqref{eq:diff1} and use the WKB approximation, we arrive at 
\bea\label{eq:WKB_result}
- 2 i [k_A(l)]^{1/2} \frac{d \tilde{A}(l)}{d l} = \epsilon \omega_p^2 \tilde{A}'_0 \exp\left[i \int^l_{l_0} (k - k_A(l')) dl'\right] \ . \nnn
\eea
Thus, the photon field induced by the DPDM can be written as
\begin{align}
\tilde{A}(l) = \tilde{A}'_0 \int^l_{l_0} dl' \frac{i\epsilon \omega_p^2(l)}{ 2 [k_A(l')]^{1/2}} 
 \exp\left[i \int^{l'}_{l_0} (k - k_A(l'')) dl''\right]
 \label{eq:converted-photon-field}
\end{align}

In the above discussions, we have used the condition \eqref{eq:WKB1}. It suggests that the alterations in $\tilde{A}$ and $\tilde{A}'$, as measured by their first and second derivatives, are significantly smaller compared to their momentum in the resonance region.
Specifically, we can make the following estimates:
\begin{align}
& | \partial_{l} \tilde{A}(l) | \approx |\tilde{A}(l)|/r_c \ll k_c |\tilde{A}(l)|, \\
& | \partial_{l}^2\tilde{A}(l)| \approx |\tilde{A}(l)|/r_c^2 \ll k_c | \partial_{l} \tilde{A}(l) |,
\end{align}
The inequalities derived above establish the validity of the WKB approximation in our case. This conclusion hinges on the smoothness of the plasma density profile in Eq.~\eqref{eq:ne_profile}. However, the presence of plasma density fluctuations at smaller scales may pose a challenge to these results. To address this issue, we have incorporated the effects of plasma density fluctuations in our analysis and demonstrated that the WKB approximation remains applicable, as detailed in the latter part of this Supplemental Material.

After establishing the validity of the WKB approximation, we revisit the calculation of the conversion probability between dark photons ($A'$) and photons. 
The field amplitude expression can be simplified using the saddle-point approximation, which states that:
\beq\label{eq:saddle-point-appro}
\left|\int dl~ g(l) {\rm e}^{-if(l)} \right| \approx g(l_c)\sqrt{\frac{2\pi}{|f''(l_c)|}}
\eeq
where we have employed the expression for the norm to eliminate the redundant phase factor. $l_c$ represents the location where $f(l)$ reaches its local maximum or minimum, corresponding to the resonance condition $\omega_p(l_c) = m_{A'}$.

As a consequence, only the second derivative plays a significant role. Its expression is given by
\beq\label{eq:second-derivative}
f''(l_c) = \left(\frac{d k_A}{d l}\right)_{l = l_c} = \left(- \frac{\omega_p}{k_A} \frac{d\omega_p}{dr}\frac{dr}{dl} \right)_{l=l_c}.
\eeq
The length of the resonant layer, denoted as $\delta l_{\rm res}$, can be estimated as the length that causes $f(l)\approx f(l_c)+  (l-l_c)^2 f''(l_c)/2$ to vary by $\pi$. This can be expressed as
\beq\label{eq:l_res}
\delta l_{\rm res} \simeq \sqrt{\frac{2\pi}{|f''(l_c)|}} \sim r_c \cdot \sqrt{\frac{\pi}{x}},
\eeq
where $x \equiv \frac{m_{A'}^2 r_c}{2 k}$. The factor $dl/dr = 1/\cos{\alpha}$ usually contributes a value of $\mathcal{O}(1)$. For $m_{A'} r_c \sim 10^7$, which is typical for the electron density distribution in space, $x \sim 10^{10}$. This large value of $x$ results in a very thin resonant layer with a thickness of $\delta l_{\rm res} / r_c \sim 10^{-5}$. This justifies the application of the saddle-point approximation, which allows us to neglect the higher-order derivatives~\cite{An:2020jmf}.

By applying the saddle-point expression of $f''(l_c)$ to Eq.~\eqref{eq:converted-photon-field}, we arrive at the simplified conversion probability, 
\begin{align}
 P_{A'\to \gamma} \approx \pi \frac{\epsilon^2 m_{A'}^4}{2 k_{A}^2} \left|\frac{\partial k_{A}(r)}{\partial l} \right|^{-1}_{l=l_c}  , 
\end{align}
which leads to Eq.~\eqref{eq:prob} in the main text after substituting $k_A$ by $\omega_p$ and $dl/v_l = dr/v_r$. The factor of $2/3$ in Eq.~\eqref{eq:prob} is included due to the inability of the longitudinal mode of a photon to propagate away. We initiated our analysis with a realistic hyperbolic trajectory and thoroughly verified that the conditions for the WKB and saddle-point approximations are met. It is noteworthy that Eq.~\eqref{eq:prob} coincides with the expression derived in Refs.~\cite{Raffelt:1987im, Mirizzi:2009iz, An:2020jmf, An:2023mvf}.


\section{Velocity distribution}

\noindent
The conversion power, Eq.~\eqref{eq:tot_power} in the main text, was derived assuming a monochromatic dark matter (DM) velocity $v_0$. However, to better reflect the physical reality, we consider the more realistic scenario where DM velocities follow a Maxwellian distribution in the Galactic frame~\cite{Drukier:1986tm,Choi:2013eda,Evans:2018bqy}.This distribution is given by:
\beq\label{eq:Maxwellian-Galaxy}
f_G(v_0) = \frac{4}{\sqrt{\pi}} \frac{v_0^2}{v_p^3}\exp\left(-\frac{v_0^2}{v_p^2}\right),
\eeq
where $v_p$ represents the most probable speed, which is taken as the speed of the Local Standard of Rest (LSR), i.e., the circular velocity around the Galactic center at the solar position. The normalization condition is $\int_{0}^{\infty} d v_0 f_G(v_0) = 1$, and we set $v_p \approx v_{\odot} \approx 220 ~\text{km}/\text{s}$. The Maxwellian distribution should be truncated at the Galactic escape speed $v_{\rm esp}$, which is $\approx 544 ~\text{km}/\text{s}$ at the solar position. However, this modification has an insignificant impact on the overall conversion rate, as $\int_{v_{\rm esp}}^{\infty} d v_0 f_G(v_0) \approx 0.66\%$, which can be safely neglected.

To determine the local dark matter velocity, we perform a Galilean boost to obtain the velocity distribution in the rest frame of the Sun (see, e.g., Ref.~\cite{Choi:2013eda}):
\begin{align}
&f_{\rm DM}(v_0)= \\
&\frac{1}{\sqrt{\pi}}\frac{v_0}{v_p v_{\odot}}
\left\{
\exp\left[-\frac{(v_0-v_{\odot})^2}{v_p^2}\right] - 
\exp\left[-\frac{(v_0+v_{\odot})^2}{v_p^2}\right]
\right\}    \nonumber
\end{align}
With the inclusion of the velocity distribution effect, the DM to photon conversion power, Eq.~\eqref{eq:tot_power}, can be averaged over the velocity profile as follows,
\beq
\mathcal{P}=\int_{0}^{\infty} dv_0 ~\mathcal{P}_0(v_0) f_{\rm DM}(v_0).
\eeq
We see that the velocity term in Eq.~\eqref{eq:tot_power} effectively cancels out with the velocity term in the denominator of the conversion probability.
Consequently, the results are not substantially influenced by the precise velocity of the dark matter particles. To further corroborate this observation,
we have numerically confirmed that the averaged power $\mathcal{P}$ is only marginally different from $\mathcal{P}_0$.

In addition, the macroscopic velocity of the plasma is typically around $400 \, \text{km/s}$ in the region between the Sun and the Earth. This velocity is comparable to the dark matter velocity, but it does not affect the calculation procedure for the conversion probability. Our focus is on the forward scattering with resonance, meaning that neither the dark photon nor the photon experiences any momentum exchange with the plasma. 
One can always transform back to an inertial frame where the plasma is at rest macroscopically, 
without changing the result.
The conversion probability exhibits a $v^{-1}$ dependence in Eq.~\eqref{eq:prob}, while the signal power is proportional to the dark matter energy flux, thus proportional to the dark matter velocity. 
Again, this shows the cancellation of velocity dependence in the total power, as discussed above, indicating that the signal strength is not sensitive to the velocity.


\section{Effect of solar magnetic field}

The influence of the solar magnetic field on conversion power appears to be marginal, as illustrated below. Measurements indicate that the magnetic field ranges from approximately 1 to 4 Gauss at distances between 1.05 and 1.35 solar radii from the center of the Sun~\cite{2020Sci...369..694Y}. In the region between the Sun and Earth, specifically at distances of 0.1 to 0.3 AU, the magnetic field is around $10^{-3}$ Gauss, diminishing to approximately $10^{-4}$ Gauss at 1 AU. In interplanetary space, which encompasses the PSP orbits (from $\sim$10 $R_{\odot}$ to 1 AU), in-situ measurements indicate that the magnetic field decreases from about 2000 nT (0.02 Gauss) near the Sun to approximately 6 nT (0.00006 Gauss) at 1 AU~\cite{Bale2016TheTransients}. 

In the magnetic field, the cyclotron frequency of the plasma is defined as $f_{\rm cycl} = eB / (2 \pi m_e)$. To estimate the effect of the magnetic field in the resonant photon-dark-photon conversion, we need to compare $f_{\rm cycl}$ to the plasma frequency, $f_{p} = \omega_{p} / (2 \pi) = (e^2 n_e / m_e)^{0.5} / (2 \pi)$. In our analysis in this work, the highest frequency is 20 MHz, corresponding to $f_{p} \approx 20~{\rm MHz}$ and resonant location $r\approx1.76~R_{\odot}$. Assuming even an aggressive estimation of the magnetic field strength of 1 Gauss, we obtain $f_{\rm cycl} = 2.7~{\rm MHz} \ll f_{p} = 20~{\rm MHz}$. As the distance from the Sun increases, $f_{\rm cycl}$ drops faster than $f_p$. 

Following~\cite{Hardy:2022ufh}, we solve the dispersion relation of the electromagnetic wave in the presence of the magnetic field. In the case that $f_{\rm cycl}\ll f_p$, the resonant condition is modified to 
\bea
m_{A'}^2\approx \omega_p^2 (1\pm \omega_{\rm cycl}/\omega_p) \ ,
\eea
where $\omega_{\rm cycl} \equiv 2\pi f_{\rm cycl}$, and the $\pm$ are for the two transverse polarizations. 


This will shift the resonant location $r_c$ and the quantity $L_c \equiv \left|d \ln{\omega_p^2(r)}/dr \right|^{-1}_{r=r_c}$ with $\Delta L_c/L_c = \Delta r_c/r_c \sim \mathcal{O}(\omega_{\rm cycl}/\omega_p)$. Since the total conversion power scales as $r_c^2L_c$, we average over the two transverse modes and obtain a modification factor in the average conversion probability,
\bea
&&\frac{1}{2}\left[(r+\Delta r_c)^2(L_c+\Delta L_c) + (r-\Delta r_c)^2(L_c-\Delta L_c) \right] \nnn
&&~~~~\sim r_c^2 L_c \mathcal{O}(\omega_{\rm cycl}^2/\omega_p^2) \ .
\eea
Therefore, the effect of the non-zero magnetic field on the signal power is at the level of $\mathcal{O}(\omega_{\rm cycl}^2/\omega_p^2) \sim 1\%$, and does not significantly affect our results.


\section{Signal detection}

\begin{figure}
    \centering
    \includegraphics[width=0.9\linewidth]{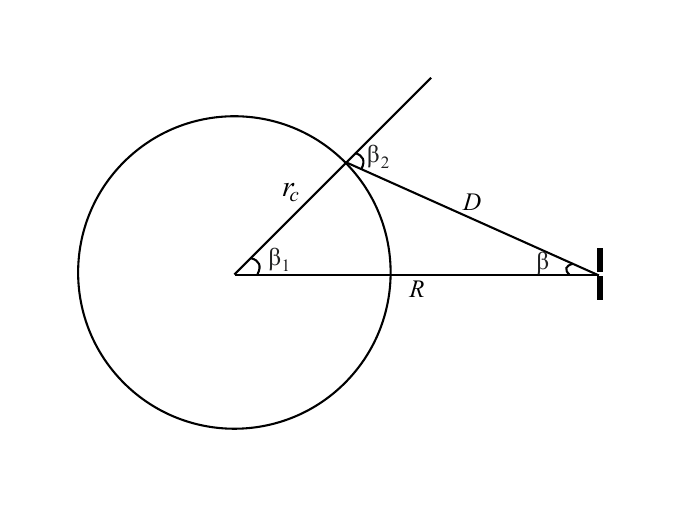}
    \caption{The sketch illustrates the geometry and defines the length and angle parameters. The spherical shell represents the resonant conversion layer. The thick black line depicts the dipole antenna of the PSP, which remains consistently perpendicular to the radial direction.}
    \label{fig:geometry}
\end{figure}

\noindent
The resonant conversion region forms a spherical shell surrounding the Sun. A solar probe intercepts the converted radio photons emanating from this spherical shell. In this section, we compute the flux density of converted radio photons that can be projected onto a solar probe equipped with a dipole antenna. 
For clarity, we illustrate the emission and detection geometry in Fig.~\ref{fig:geometry}, with the relevant lengths and angles labeled.
We plot the dipole antenna perpendicular to the radial direction, consistent with the orientation of the PSP antenna plane, which is always maintained perpendicular to the radial direction~\cite{Bale2016TheTransients, Pulupa2017TheProcessing}.

First, we start with the (spectral) flux density emitting out from the spherical conversion layer, which is
\beq
S_{c,0} = \frac{1}{4\pi r_c^2} \frac{1}{\mathcal{B}}\mathcal{P}.
\eeq
At a specific location on the conversion layer, the emission typically follows an angular distribution $f_c$ relative to the polar angle $\beta_2$ (refer to Fig.~\ref{fig:geometry}). This distribution can be represented by
\beq
S_c(\beta_2) = S_{c,0}  f_c(\beta_2). 
\eeq
where the angular distribution $f_c(\beta_2)$ is normalized as 
\beq\label{eq:f_normalization}
\int_{0}^{1}  d\cos\beta_2 ~ f_c(\beta_2) =\frac{1}{2\pi}.
\eeq
Please note that the flux at any given point on the conversion sphere radiates outward into a $2\pi$ space rather than the entire $4\pi$ space. Although the DPDM originates from all directions in the $4\pi$ space, the converted photons traveling inward will experience complete reflection due to the denser plasma in the inner region~\cite{An:2020jmf}.

Next, similar to Refs.~\cite{An:2023wij, favorite2016solid}, we can calculate the flux received by the dipole antenna using the following expression:
\beq
S' &= 
\int_{0}^{2\pi} d\phi \int_{\frac{r_c}{R}}^{1} d\cos\beta_1 ~
 \frac{r_c^2 S_{c,0} f_c(\beta_2)}{D^2}
\frac{1-\sin^2{\beta}\sin^2{\phi}}{2}.
\label{eq:flux_at_R_0}
\eeq
Here, we integrate over the polar angle $\beta_1$, with the Sun as the center, ranging from $0$ to $\arccos(r_c/R)$, and over the azimuthal angle $\phi$. It's important to note that, unlike in Ref.~\cite{favorite2016solid}, which discusses a point detector collecting particles, here we consider a dipole antenna collecting radio waves. Therefore, we need to account for the projection of the electric field $E$ (specifically, $E^2$ since we are discussing energy flux) onto the dipole antenna. 

The light propagates a distance $D$ to arrive at the dipole antenna. The polarization of the electric field $E$ is perpendicular to the momentum of light. Therefore, the projection of $E^2$ onto the dipole is given by:
\begin{align}
E_{\parallel}^2 = E^2 \frac{1-\sin^2\beta\sin^2\phi}{2}.
\end{align}
This explains why we replace the simple $\cos\beta$ term in Ref.~\cite{favorite2016solid} with $(1-\sin^2\beta\sin^2\phi)/2$ in Eq.~\eqref{eq:flux_at_R_0}. The factor $1/2$ is due to averaging over the light polarizations.

Using the following geometric relations (refer to Fig.~\ref{fig:geometry}):
\beq
&D = \sqrt{R^2 + r_c^2 - 2Rr_c\cos\beta_1}, \\
&\cos\beta_2 = \frac{R\cos\beta_1 - r_c}{D}, \\
& \frac{d\cos\beta_2}{d\cos\beta_1} = R^2\cdot \frac{R-r_c\cos\beta_1}{D^3},\\
& \cos\beta = \frac{R-r_c\cos\beta_1}{D},
\eeq
we can rewrite Eq.~\eqref{eq:flux_at_R_0} as
\begin{align}
S' &= 
\frac{r_c^2}{2R^2}\int_{0}^{2\pi} d\phi \int_{0}^{1} d\cos\beta_2 ~
 S_{c,0} f_c(\beta_2)
\frac{1-\sin^2{\beta}\sin^2{\phi}}{\cos\beta} \nonumber \\
&= 
\frac{\pi S_{c,0} r_c^2}{R^2}\int_{0}^{1} d\cos\beta_2 ~
  f_c(\beta_2)
\frac{1+\cos^2{\beta}}{2\cos\beta}. 
\label{eq:flux_at_R}
\end{align}
Since $1+\cos^2{\beta} \geq 2\cos\beta$ and $0\leq \beta <\pi/2$, we always have $(1+\cos^2{\beta})/(2\cos\beta) \geq 1$. Thus, the integral in the second equality of Eq.~\eqref{eq:flux_at_R} is always not smaller than Eq.~\eqref{eq:f_normalization}, yielding
\beq
S' \geq \frac{S_{c,0} r_c^2}{2R^2} 
= S_{\rm sig}.
\eeq
Recall that $S_{\rm sig}$ has been defined in Eq.~\eqref{eq:S_sig} in the main text. When the flux angular distribution is a Dirac delta function, $f_{c}(\beta_2) = \delta(\beta_2) /(2\pi\sin\beta)$, we have $S' =  S_{\rm sig} = 1/2 \cdot 1/(4\pi R^2) \cdot \mathcal{P}/\mathcal{B}$ as expected, which reaches its minimum. If the function $f_{c}(\beta_2)$ takes other forms, $S'$ is generally larger than $S_{\rm sig}$. However, as the satellite's location $R$ increases away from $r_c$, $S'$ rapidly becomes identical to $S_{\rm sig}$. For example, if the local converted photon emits as a spherical uniform distribution, $f_c(\beta_2) = 1/(2\pi)$, then solving Eq.~\eqref{eq:flux_at_R} gives:
\beq
\frac{S'}{S_{\rm sig}} = 
\frac{1}{8}\left[2+ \left(\frac{1}{x} - 3x\right)\ln\left( \frac{x-1}{x+1}\right)\right]
,~
x\equiv \frac{R}{r_c}.
\eeq
At $x=1$, the ratio $S'/S_{\rm sig}$ becomes positively divergent. However, this divergence should not be interpreted as physical, but rather attributed to the improper neglect of the finite size of the dipole antenna at $R-r_c = 0$. As $x$ increases, the dipole size indeed becomes negligible compared with $R-r_c$, and the ratio $S'/S_{\rm sig}$ rapidly drops towards unity.

More realistically, the angular distribution function $f_{c}(\beta_2)$ would tend to concentrate along the center line $\beta_2 = 0$. This occurs due to the refraction effect: after leaving the conversion layer, the converted photons are efficiently refracted to travel along the radial direction due to the decreasing plasma density~\cite{An:2020jmf}. 
The refraction law is $n\sin{\theta} = \text{constant}$, where $n$ is the refractive index and $\theta$ is the photon momentum angle relative to the radial direction $\hat{r}$.
The refraction index is given by $n(r, \omega) = [1-\omega_p^2(r)/\omega^2]^{1/2}$ where $\omega_p$ is the plasma frequency and $\omega$ is the photon energy. At the resonant layer, $\omega_p(r_c) = m_{A'}$ and $\omega^2 = m_{A'}^2(1+v_{A'}^2)$, and the corresponding refractive index is thus $n(r_c,\omega) \approx v_{A'} \approx 10^{-3}c$. The refraction effect is highly efficient. To demonstrate this, we consider a small decrease in $\omega_p(r)$. For example, if $\omega_p$ decreases by $10^{-4}$ of $m_{A}'$ at the radius $r_1$, that is, $\omega_{p}(r_1) = (1-10^{-4})m_{A'}$, it results in the refractive index $n(r_1,\omega) \approx 0.014$. This constrains the photon emission angle $\sin\theta(r_1) \approx (v_{A'}/0.014)\sin\theta(r_c) \leq v_{A'}/0.014 \approx 7\times 10^{-2}$ and thus $\theta(r_1) \leq 4.1^{\circ}$.
Although this refraction effect may not be as effective for in-situ measurements due to the relatively short distance between the conversion layer and the probe's position, the exact form of $f_{c}(\beta_2)$ is less critical in this context. To be conservative, we will consider $S' = S_{\rm sig}$ as the (spectral) flux density detectable by a dipole antenna.

\begin{figure}
    \centering
\includegraphics[width=0.8\linewidth]{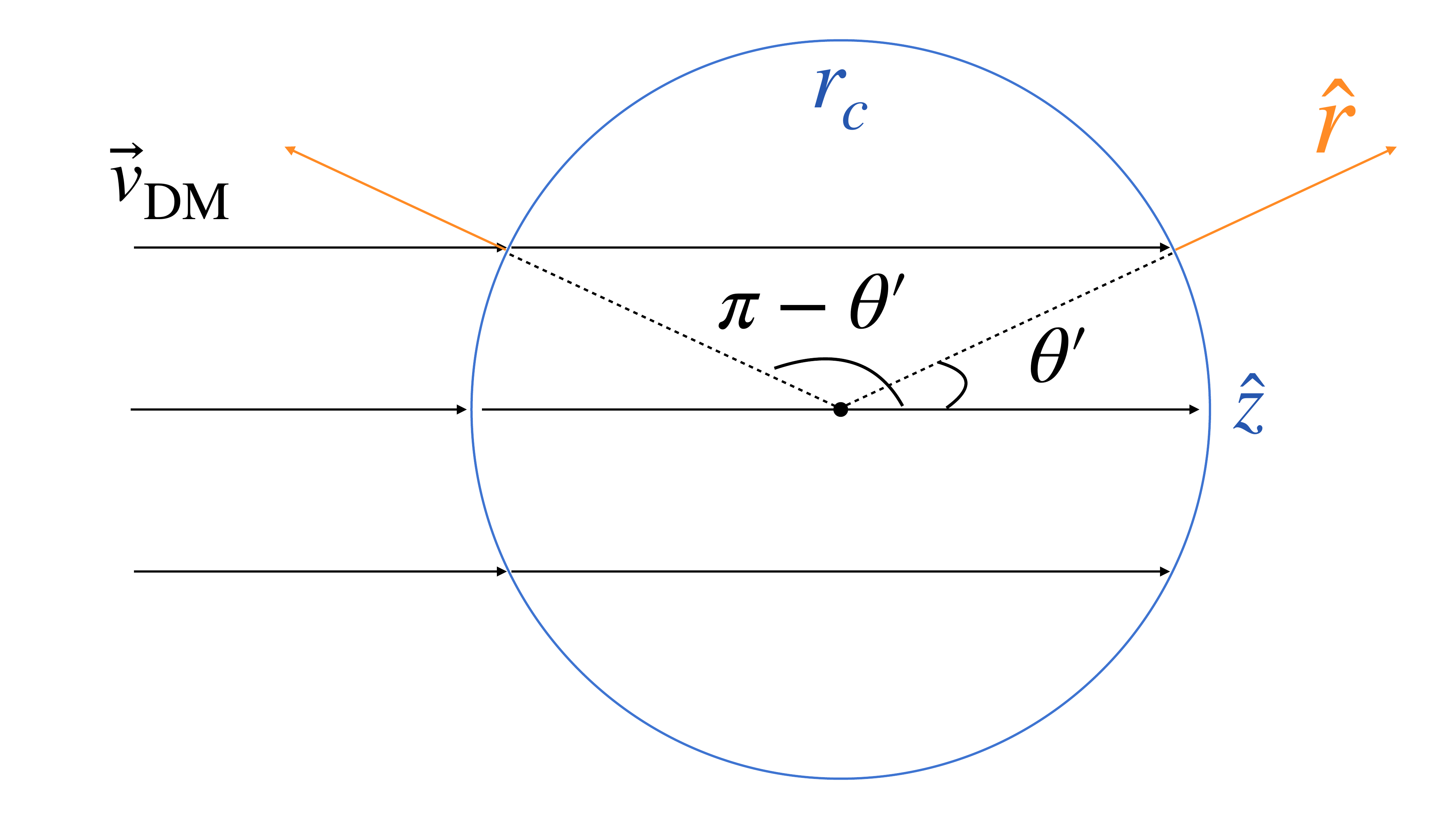}
    \caption{
    Illustration of incident dark matter (black arrow) with fixed velocity \( \vec{v}_{\rm DM} \) and the radial emission of photons (orange arrow) in the direction \( \hat{r} \). 
    }
    \label{fig:refraction-illustration}
\end{figure}

Next, we discuss the effect of anisotropy of dark matter velocity distribution on detection, which is induced by the relative motion between the Sun and the DM halo. It turns out that even with the anisotropy in the DM velocity distribution, the emission of converted photons can still be regarded as spherically symmetric. 
Assume the incoming DM stream has a fixed velocity \( \vec{v} \) along the \( z \)-direction. The power of emitted photons per unit area in a given direction \( \hat{r} \) is:  
$\text{Power}(\hat{r}) = \rho_{\rm DM} \times |\vec{v} \cdot \hat{r}| \times P_{A' \to \gamma}(v_r)$,
where the second term, \( |\vec{v} \cdot \hat{r}| \), accounts for the projection of the unit area perpendicular to \( \vec{v} \). On the far side of the resonant layer, photons are emitted in equal amounts due to the negligible attenuation of the DM flux, as illustrated in Fig.~\ref{fig:refraction-illustration}. 
Recalling \( P_{A' \to \gamma}(v_r) = c_0 v_r^{-1} \) from Eq.~\eqref{eq:prob} in the main text, where \( c_0 \) is constant and \( v_r = |\vec{v} \cdot \hat{r}| \), the power becomes 
$\text{Power}(\hat{r}) = \rho_{\rm DM} c_0$,
which is independent of the radial direction \( \hat{r} \). This result arises because the reduced projection area is precisely compensated by the enhanced conversion probability from the lower radial velocity.


\section{Data calibration}

\noindent
The radio-frequency signals measured by a space probe originate from various sources. The Galactic radio background dominates the spectrum above approximately 500 kHz. Below a few hundred kHz, the spectrum is dominated by space plasma radio sources, which include electron quasi-thermal noise, electron shot noise, proton noise, and other sources. Additionally, instrumental noise generated by the probe itself contributes as well,

Space probes detect radio-frequency signals by monitoring the fluctuations in electric potential at the antenna terminals. The strength of the signal is typically quantified as the time-averaged voltage power spectral density, denoted as $V^2$ and measured, for instance, in units of ${\rm V}^2 \, {\rm Hz}^{-1}$~\cite{zaslavsky2011antenna, Pulupa2017TheProcessing}. This parameter is related to the incident spectral flux density ($S$), measured in units of ${\rm W} {\rm m}^{-2} {\rm Hz}^{-1}$, through the following relation~\cite{manning2000instrumentation, zarka2004jupiter, eastwood2009measurements, zaslavsky2011antenna}
\beq 
V^2 = (\Gamma l_{\rm eff})^2 \cdot S,
\label{eq:V2_original}
\eeq
where we have adopted the natural unit system. Here, $l_{\rm eff}$ represents the effective length of the dipole antenna, which characterizes the electrical response of the antenna.
The gain factor $\Gamma$ is given by:
\beq
\Gamma^2 = \left| \frac{Z_s}{Z_a+Z_s}\right|^2 ,
\label{eq:Gamma}
\eeq 
where $Z_a$ is the impedance of the dipole antenna. 
The antenna is connected to the receiver through various electrical components and cables with an impedance $Z_s$, commonly referred to as the stray impedance.

The data $V^2$, measured in unit of ${\rm V}^2 \, {\rm Hz}^{-1}$, is referred to as Level-2 data. Using the relationship given by Eq.~\eqref{eq:V2_original}, $V^2$ measured by PSP has been calibrated into the spectral flux density $S$, expressed in unit of ${\rm W} \, {\rm m}^{-2} \, {\rm Hz}^{-1}$. This calibrated result of PSP, referred to as Level-3 data, is publicly accessible on the website~\cite{PSPdataLevel3}. 
It is noteworthy that the instrumental background noise has been removed from the Level-3 data.

Similarly, the DPDM-induced signal is recorded as 
\beq\label{eq:V2-S_DPDM}
V_{\rm sig}^2 = (\Gamma l_{\rm eff})^2 \cdot S_{\rm sig}.
\eeq 
Note that this expression is conservative, as we have equated the induced signal $S'$ with $S_{\rm sig}$, as discussed in the previous section.

It is sufficient to use Level-3 data to search for the DPDM signal, $S_{\rm sig}$. However, for completeness, we provide further details on how to calibrate Level-2 data into Level-3 data below. Although this additional information may not be necessary due to the availability of PSP Level-3 data online~\cite{PSPdataLevel3}, it offers a cross-check as it enables users to directly calibrate the Level-2 data. Additionally, the same calibration method can be applied to the STEREO case.

In the short-dipole regime, where the physical length of a dipole antenna $l_a$ (tip to tip, which is $7$~m for PSP~\cite{Moncuquet2020First/FIELDS} and $10.4$~m for STEREO~\cite{zaslavsky2011antenna} respectively) and the electromagnetic wavelength $\lambda$ satisfy $l_a<\lambda/10$, both $\Gamma$ and $l_{\rm eff}$ are constant with respect to the frequency~\cite{zaslavsky2011antenna, balanis2015antenna}. 
Ref.~\cite{zaslavsky2011antenna} obtained $\Gamma l_{\rm eff}\approx 2.04 \, \text{m}$ for STEREO under the short-dipole approximation by fitting the data with the Galactic radio background. Following the same method, Ref.~\cite{Maksimovic2020AnticorrelationHelios} obtained $\Gamma l_{\rm eff}\approx 1.17 \, \text{m}$ for PSP. 

As the frequency gradually increases from the short-dipole regime to the half-wavelength dipole case ($l_a=\lambda/2$), we do not have simple expressions for $\Gamma$ and $l_{\rm eff}$ as functions of frequency. However, we expect both $\Gamma$ and $l_{\rm eff}$ to increase with frequency from $l_a<\lambda/10$ to $l_a=\lambda/2$. 
The $l_{\rm eff}$ becomes larger from the value of $l_a/2$ in the short-dipole case to $2l_a/\pi$ in the have-wavelength case~\cite{balanis2015antenna}. $\Gamma$ also increases as the configuration leaves the capacitive regime in the short-dipole case towards the electric circuit resonance at the half-wavelength case~\cite{zaslavsky2011antenna}. Moreover, Ref.~\cite{zaslavsky2011antenna} has clearly shown an increase by $\mathcal{O}(10)$ of the combined factor $\Gamma l_{\rm eff}$ based on the analysis of observation data. 

Both PSP and STEREO measure the frequency below about the half-wavelength case. Therefore, we can apply the value of $\Gamma l_{\rm eff}$ for the short dipole case, \ie $\Gamma l_{\rm eff}\approx 1.17$m (PSP) and $2.04$m (STEREO), to the whole observation frequency range while placing ourselves on the conservative side at the same time. Based on the value of $\Gamma l_{\rm eff}$ and Eq.~\eqref{eq:V2_original}, we have calibrated PSP data, which turns out to agree well with the published Level-3 data~\cite{PSPdataLevel3}. Similarly, we calibrate the STEREO data~\cite{SQTN_WEB} and will use it for further data analysis. For PSP, we will instead use the Level-3 data published online~\cite{PSPdataLevel3} because it offers the advantage of removing instrumental background noise.


\section{PSP orbits and the $n_e$ profiles}

\noindent
As described in the main text, a PSP phase represents one complete orbit, encompassing two consecutive aphelion dates as its starting and ending points. The phases, along with their corresponding data ranges, the starting aphelion date and perihelion distances from the solar center, are summarized in Table~\ref{Table:PhaseNO}. 

It is worth noting that we exclude the 8th and 11th encounter phases from our analysis due to their substandard data quality \cite{PSP_FIELDS_Release_Note}. 
During the outbound phase of PSP Encounter 8, intermittent noise was observed in electric field measurements (both AC- and DC-coupled) obtained by the RFS, the Time Domain Sampler, and Digital Fields Board receivers. The RFS is sensitive to the weakest electric fields among the FIELDS/PSP receivers and therefore was more strongly contaminated. 
In the worst situation, the antenna voltage even approached the instrument measurement limit at around $-100$ V. This voltage saturation issue primarily affected antennas V1 and V2, while V3 and V4 continued to operate nominally. This troublesome condition persisted until new FIELDS Antenna Electronics Board (AEB) settings were uploaded to effectively restore the instrument to nominal operations during a command pass on June 30th, 2021.

For the phase 11,
a data gap in the RFS/LFR data, lasting around 4 days near the PSP orbit perihelion, is caused by a command error. Fortunately, RFS/HFR data was recorded normally except for during bias sweeps. Since the frequency of the local plasma line falls in the measurement range of RFS/LFR, it is impossible to derive any electron density/temperature measurements via the quasi-thermal noise (QTN) technique during the data gap. This command issue causing the RFS/LFR data gap was identified and resolved from  
phase 12, 
and therefore the instrument operated normally after that. Therefore, we exclude the 
phases 8 and 11
in our further analysis.

Onboard PSP, there are several ways to measure the in situ plasma density including the QTN technique~\cite{2017meyervernet} and the Solar Wind Electrons, Alphas, and Protons (SWEAP) instrument suite \cite{Kasper2016}. The QTN technique provides precise electron density and estimated electron temperature measurements in the solar wind \cite{2020Moncuquet,Maksimovic2020,2022Martinovic,2023Liu_QTN,Zheng_2024}. The SWEAP 
instrument suite onboard PSP comprises the Solar Probe Cup (SPC) and the Solar Probe Analyzers (SPAN) \cite{Case2020}. The SPC is a fast Faraday cup designed to measure the one-dimensional velocity distribution function (VDF) of ions. SPAN, on the other hand, is a combination of three electrostatic analyzers that operate to measure the three-dimensional ion and electron VDFs \cite{2022Livi, Whittlesey_2020}. QTN-derived electron density measurements, obtained from spectral peaks, are independent of gain calibrations, providing more reliable and accurate results. As a result, electron number density derived from QTN spectroscopy is considered the "gold standard" for density measurements and is routinely used to calibrate other instruments. On PSP, the electron number density provided by the QTN technique has played a pivotal role as a calibration standard for scientific analysis \cite{2021Kasper,2021Zhao,2021Liu,2021ApJLiu,2024LiuYingD}. As a result, we use the QTN-derived electron density measurements for analysis, and they are available in ~\cite{Moncuquet2020First/FIELDS, SQTN_WEB}. 

While PSP provides valuable electron density ($n_e$) measurements within 0.5 AU of the Sun, it faces limitations when extending these measurements to larger distances due to the Debye length becoming much larger than the dipole length of PSP antenna \cite{2017meyervernet}. This restricts the PSP $n_e$ measurements to a maximum of $\pm 15$ days from each perihelion passage~\cite{2020Moncuquet}. To address this limitation and obtain a comprehensive $n_e$ profile across the solar-wind plasma, we utilize the in-situ $n_e$ data to fit the approximated formula~\eqref{eq:ne_profile}. This enables us to extrapolate the $n_e$ profile to 1 AU. We perform this fitting for each PSP phase, and the corresponding $n_e(1$ AU$)$ values are summarized in Table~\ref{Table:PhaseNO}. This approach provides an explicit solar-wind electron density profile for each PSP phase. When calculating the DPDM-induced signal and comparing it to PSP data collected on a specific day, we employ the $n_e$ profile corresponding to the phase within which that day lies. This strategy effectively mitigates uncertainties arising from $n_e$ fluctuations.

For STEREO observations, we employ the reference electron density value of $n_e(1{\rm~AU}) = 7.2~{\rm cm}^{-3}$ from Ref.~\cite{1998SoPh..183..165L}. This value serves as a benchmark for comparison with the PSP results. While the electron density at 1~AU for STEREO can fluctuate around $\sim 10~{\rm cm}^{-3}$, these fluctuations have a minor impact on the final STEREO constraints.

\begin{table}[!htp]
    \centering
    \begin{tabular}{c|c|c|c}
    \hline
    \hline
            \begin{tabular}{@{}c@{}}  Phase \\  No. \end{tabular} 
          & \begin{tabular}{@{}c@{}}  Date Range \\ (yy/mm/dd) \\ (UTC time)   \end{tabular} 
          & \begin{tabular}{@{}c@{}}  Aphelion/Perihelion \\ ($R_{\odot}$) \\ (UTC time 12:00:00)  \end{tabular}  
          & \begin{tabular}{@{}c@{}}  $n_e$ at $1$~AU  \\ (cm$^{-3}$)  \end{tabular} \\
          \hline
          1  &  \begin{tabular}{@{}c@{}} 18/10/02 \footnote{The first date with data available rather than an aphelion date. 
          }
          \\ 19/01/19 \end{tabular} 
             & ~207.82/35.78  & 11.0 \\
             \hline
          2  &  \begin{tabular}{@{}c@{}} 19/01/20 \\ 19/06/17  \end{tabular}    
             & ~201.72/35.83  & 8.7 \\
             \hline
          3  & \begin{tabular}{@{}c@{}} 19/06/18  \\ 19/11/14  \end{tabular}    
             & ~201.72/35.72  & 8.0 \\
             \hline
          4  & \begin{tabular}{@{}c@{}} 19/11/15  \\ 20/04/02  \end{tabular}    
             & ~201.72/27.89  & 12.3 \\
             \hline
          5  & \begin{tabular}{@{}c@{}} 20/04/03  \\ 20/08/01  \end{tabular}    
             & ~188.06/27.91  & 14.1 \\
             \hline
          6  & \begin{tabular}{@{}c@{}} 20/08/02  \\ 20/11/21  \end{tabular}    
             & ~175.61/20.39  & 10.0 \\
             \hline
          7  & \begin{tabular}{@{}c@{}} 20/11/22  \\ 21/03/08  \end{tabular}        & ~175.61/20.51  & 13.6 \\
             \hline
          9  & \begin{tabular}{@{}c@{}} 21/06/19  \\ 21/09/29  \end{tabular}    
             & ~168.28/16.39  & 14.9 \\
             \hline
          10 & \begin{tabular}{@{}c@{}} 21/09/30  \\ 22/01/07  \end{tabular}        & ~168.27/13.43  & 7.9 \\
             \hline
          12 & \begin{tabular}{@{}c@{}} 22/04/14  \\ 22/07/19  \end{tabular}        & ~163.55/14.63  & 9.1 \\
    \hline
    \hline
    \end{tabular}
    \caption{The detailed information for each PSP phase. In creating this table, we relied on the PSP position calculator website~\cite{PSPwebPosition}.}
    \label{Table:PhaseNO}
\end{table}


\section{Data processing}

\noindent
A comprehensive dataset of PSP radio data spanning from September 2, 2018, to July 19, 2022, is utilized in our analysis. We use the Level-3 data publicly available online~\cite{PSPdataLevel3}, which has already been calibrated to the spectral flux density in unit of ${\rm W} {\rm m}^{-2} {\rm Hz}^{-1}$ based on the relation Eq.~\eqref{eq:V2_original}.
Within each day, the background data is determined by selecting the lowest $3\%$ in observed values of data points. This filtering approach is widely employed in data processing for satellites like PSP and STEREO~\cite{zaslavsky2011antenna, Maksimovic2020AnticorrelationHelios}. It effectively removes large data fluctuations, such as radio bursts, while preserving the DPDM-induced signal, which remains constant in the spectrum in short time period of a day.

The LFR and HFR instruments onboard PSP each have 64 frequency bins. Each frequency bin contains a varying number of data points in the time series, with the exact number varying from day to day. For each frequency bin $i$, we select the lowest $3\%$ of data points and label them as $O_{i,j}$, where $j$ represents the time bins and $N_i$ is the total number of data points in that frequency bin. The filtered data, representing the background of that day and any possible DPDM-induced signal, is then averaged to obtain the average $\bar{O}_i$ for that frequency bin. Therefore, the average $\bar{O}_i$ and the standard error $\sigma_i$ for the average are calculated as
\begin{align}
\bar{O}_i & = \frac{1}{N_{i}} \sum_{j=1}^{N_i} O_{i,j} , \\
\sigma_i^2 & = \frac{1}{N_i(N_i-1)} \sum_{j=1}^{N_i} (O_{i,j} - \bar{O}_i)^2 ,  
\end{align}
respectively. It is crucial to note that if the number of data points $N_i$ for a specific frequency bin falls below a minimum threshold of 10, we will exclude the observation for that day from further analysis. This stringent criterion ensures that the data points used for analysis are sufficiently numerous and representative of the statistics.

To further refine the background estimation, we employ a polynomial function to fit the data, together with a negative-power term
\beq
\label{eq:bkg_func}
B(a, f) = \sum_{q=0}^{n} {a_q} f^q + a_{-3} f^{-3}
\eeq
where $f$ represents frequency and $a=\{a_q, a_{-3}\}$ are the coefficients. This fitting function locally approximates $\bar{O}_{i_0}$ and its neighboring frequency bins from $i = i_0 - k$ to $i = i_0 + k$. In practice, we choose $n=3$ and $k=5$. 
The term $a_{-3} f^{-3}$ accounts for the behavior of quasi-thermal noise (QTN), which is one of the main sources of radio signals at low frequencies (see \eg Ref.~\cite{zaslavsky2011antenna}). It has a $f^{-3}$ dependence on frequency as described in Refs.~\cite{zaslavsky2011antenna, 2017meyervernet, meyer1989tool} 

\begin{figure}
    \centering
\includegraphics[width=1\linewidth]{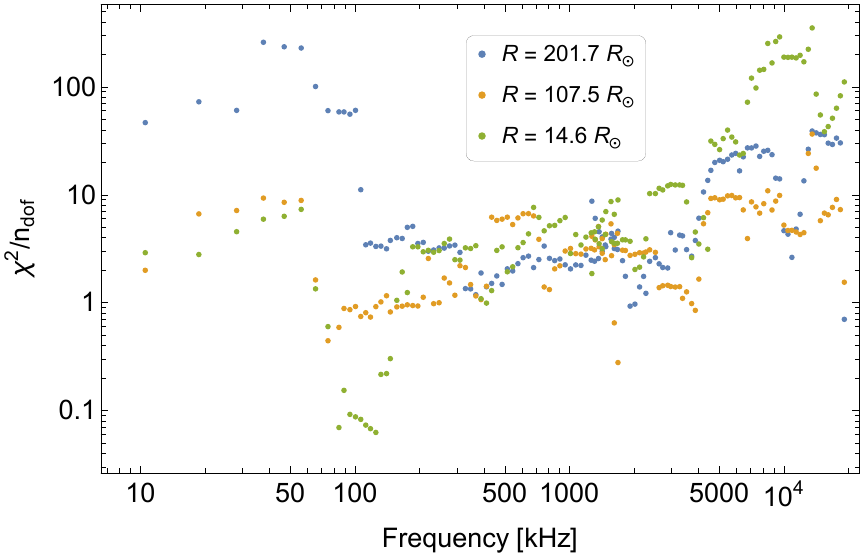}
\caption{    The values of $\chi^2/n_{\rm dof}$. We randomly choose three days as examples to show the results. They are at different distances from the Sun, $201.7~R_{\odot}$ (November 15, 2019), $107.5~R_{\odot}$ (January 01, 2021) and $14.6~R_{\odot}$ (June 01, 2022).}
\label{fig:chisquaredOverNdof}
\end{figure}

To avoid underestimating the errors, we follow the method in Ref.~\cite{ParticleDataGroup:2018ovx}. Let $\tilde{a}$ denote the optimal set of coefficients that minimize the weighted least squares error
\beq\label{eq:chi_rescaling}
\chi^2 = \sum_{i=i_0-k}^{i_0+k} \left[ \frac{B(\tilde{a}, f_i) - \bar{O}_i}{\sigma_i} \right]^2.
\eeq
The reduced $\chi^2$ is defined as $\chi^2/n_{\rm dof}$ where $n_{\rm dof}=(2k+1)-(n+2)$ is the effective number of degrees of freedom. If the reduced $\chi^2$ is larger than unity, then all errors $\sigma_i$ should be rescaled by multiplying a common factor, $\tilde{\sigma}_i = \sigma_i\sqrt{\chi^2/n_{\rm dof}}$.
As we are going to see later in the section Robustness of the background fitting, this method of error rescaling is consistent with the error estimation method used in Refs.~\cite{An:2022hhb, An:2023wij}. 
In Fig.~\ref{fig:chisquaredOverNdof}, we present $\chi^2/n_{\rm dof}$ using PSP datasets from several arbitrarily selected days as illustrative examples.  

Similarly, we have applied the above process to the STEREO data. We utilized the STEREO data observed on January 13, 2007, the day on which the Sun is relatively quiet~\cite{zaslavsky2011antenna}. We have selected the lowest $1\%$ of data as the background of that day, the same as that used in Ref.~\cite{zaslavsky2011antenna}, which can ensure enough data points to represent the statistics.


\section{Statistical test}

\noindent 
To establish constraints on the DPDM signal strength and, in turn, the DPDM-photon coupling strength $\epsilon$, we employ a likelihood-based method for data analysis~\cite{Cowan2011, An:2022hhb, An:2023wij, An:2024kls}. Initially, we construct a likelihood function at the frequency bin $i_0$ along with its neighboring $2k$ bins, 
\begin{align}
  &   L(S_{\rm sig},a) =  \label{eq:likelihood} \\
& \prod_{i=i_0-k}^{i_0+k} \frac{1}{\sqrt{2\pi}\tilde{\sigma}_i} \exp{\left\{
-\frac{1}{2} \left[
\frac{B(a,f_i)+ S_{\rm sig}\delta_{ii_0} - \bar{O}_i}{\tilde{\sigma}_i}
\right]^2
\right\}}, \nonumber
\end{align}
where $\bar{O}_i$ represents the filtered data with its corresponding rescaled uncertainty $\tilde{\sigma}_{i}$ as discussed in the previous section. 
$B(a, f)$ represents the function~\eqref{eq:bkg_func} that locally fits the background data
at the neighboring frequency bins $f_i$, where the polynomial coefficients $a$ and $a_{-3}$ are treated as nuisance parameters. A DPDM-induced signal $S_{\rm sig}$ is assumed to exist at the frequency bin $i_0$. 
Such a signal is monochromatic because the dark matter bandwidth is 
\beq
\mathcal{B}_{\rm DM}\sim
\frac{m_{A'}v_0^2}{2\pi} \sim 
6\times 10^{-4}~{\rm kHz} \times \left( \frac{m_{A'}/2\pi}{{\rm MHz}}\right) ,
\eeq
which is much narrower than the frequency resolution of PSP and STEREO.

To evaluate the statistical significance of the DPDM signal, we employ a likelihood-based test statistic $q_S$, defined as:
\beq 
q_S  =
\begin{cases}
-2\ln{\left[\frac{L(S_{\rm sig},\hat{\hat{a}})}{L(\hat{S}_{\rm sig},\hat{a})}\right]}, & \hat{S}_{\rm sig}\leq S_{\rm sig} \\
0, & \hat{S}_{\rm sig}> S_{\rm sig}
\end{cases},
\eeq
where $L(S_{\rm sig}, a)$ represents the likelihood function, $\hat{S}_{\rm sig}$ and $\hat{a}$ are the values of $S_{\rm sig}$ and $a$ that maximize the likelihood function, and $\hat{\hat{a}}$ is the value of $a$ that maximizes $L(S_{\rm sig}, a)$ for a particular $S_{\rm sig}$. 

The test statistic $q_S$, which is a function of $S_{\rm sig}$, follows a half-chi-square distribution~\cite{Cowan2011}
\beq
f(q_S|S_{\rm sig}) = \frac{1}{2}\delta(q_S)+\frac{1}{2}\frac{1}{\sqrt{2\pi}}\frac{1}{\sqrt{q_S}}{\rm e}^{-q_S/2}.
\eeq
Its corresponding cumulative distribution is denoted as $\Phi(\sqrt{q_S})$ where the function $\Phi(x)$ is the unit normal distribution (the mean is $1$ and the standard deviation is $0$).
To quantify the statistical significance of the assumed signal $S_{\rm sig}$, we define the $p$-value as
\beq
p_S = \frac{1-\Phi(\sqrt{q_S})}{1-\Phi(\sqrt{q_0})},
\eeq
where $q_S$ represents the test statistic calculated under the assumption of a non-zero signal $S_{\rm sig}$, and $q_0$ is the test statistic obtained by setting $S_{\rm sig} = 0$. It allows us to compare the strength of the assumed signal to the null hypothesis of no signal and determine its statistical significance.
Practically, we set $p_S = 0.05$ to derive the signal flux $S_{\rm sig}$, which serves as the upper bound on the signal flux $S_{\rm lim}$, indicating a rejection of the signal at the $95\%$ confidence level. The constraint on $S_{\rm sig}$ ($S_{\rm sig}  < S_{\rm lim}$) is then translated into a constraint on the kinetic mixing constant $\epsilon$ using Eq.~\eqref{eq:S_sig} in the main text.


\section{Robustness of the background fitting}

\begin{figure}
    \centering
    \includegraphics[width=1\linewidth]{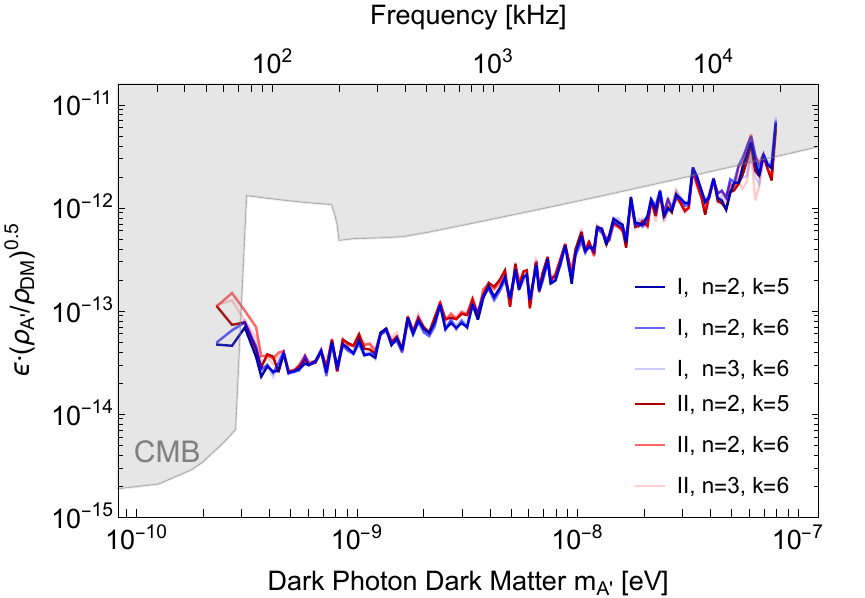}
    \caption{Comparisons of the upper limits derived from different choices of parameters ($n$ and $k$) and different methods of assessing uncertainties (method I and II). Additionally, the existing constraints from CMB observations~\cite{Arias:2012az, Witte:2020rvb} are shown as the shaded gray region. 
    This is plotted with the date December 01, 2018 as an example.
    }
    \label{fig:robust_comparison}
\end{figure}

\noindent
The log-likelihood based statistic test sets the upper limits on the monochromatic signal and thus $\epsilon$. This process turns out to be insensitive to background-fitting parameters, $n$, the degree of the polynomial fitting function, and $2k+1$, the number of bins in a fitting. This has been explicitly demonstrated in Ref.~\cite{An:2023wij}. Such insensitivity also holds in our case. To see this, we present in Fig.~\ref{fig:robust_comparison} the upper limits on $\epsilon$ obtained with different choices of fitting parameters: $(n,k) = (2,5), (2,6)$ and $(3,6)$ respectively, plus the negative-power term $a_{-3} f^{-3}$. The results exhibit remarkable similarity especially within the parameter space of interest, highlighting the robustness of the constraints against variations in parameters.

In addition, we present an alternative method of assessing the uncertainties introduced by the fitting function~\cite{An:2022hhb, An:2023wij}. We refer to the error-rescaling based on Eq.~\eqref{eq:chi_rescaling} as method I, while we denote the approach described below as method II. 
The deviations of the data points from the fit function are treated as systematic uncertainties, which can be calculated as:
\beq
\left( \sigma_{i_0}^{\rm sys} \right)^2 = \frac{1}{2k-1}\sum_{i=i_0-k}^{i_0+k}(\delta_i - \bar{\delta})^2
,
\text{($i = i_0$ excluded)}.
\eeq
$\delta_i \equiv  B(\tilde{a}, f_i) - \bar{O}_i$ where $\bar{\delta}$ is the average of the $\delta_i$ values and $\tilde{a}$ represents the optimal set of coefficients that minimize the weighted least squares error ($i_0$ excluded).
The total uncertainty at frequency bin $i_0$ is then calculated by taking the square root of the sum of the statistical and systematic uncertainties, $\sigma_{i_0}^{\rm tot} =
\sqrt{
\sigma_{i_0}^2 
+
\left( \sigma_{i_0}^{\rm sys} \right)^2
}.$
Replacing the rescaled uncertainty $\tilde{\sigma}_i$ in the likelihood Eq.~\eqref{eq:likelihood} by this total uncertainty $\sigma_{i}^{\rm tot}$, we then repeat the process of calculating the upper limits. The results based on this method II of evaluating uncertainties are shown in Fig.~\ref{fig:robust_comparison} for different choices of $n$ and $k$, together with the results from the method I. The similarity between these results obtained using the two methods once again demonstrates the robustness of the upper limits. 

\section{Effects of plasma density fluctuations}

\noindent
In our calculations, the solar-wind plasma is treated as a steady flow from the solar corona to 1 AU. This assumption is well-established, particularly during solar minimum periods; see \eg Ref.~\cite{1998SoPh..183..165L}. The observation dates selected for PSP (2018-2022) and STEREO (2007) in this study fall around two consecutive solar minima. Consequently, we have adopted Eq.~\eqref{eq:ne_profile} as the electron density profile. However, it is worth noting that the steady flow assumption may be challenged by density fluctuations. These fluctuations could potentially undermine the validity of the conversion probability formula, Eq.~\eqref{eq:prob}, derived in the section Conversion Probability in this Supplemental Material. Additionally, these fluctuations could introduce non-spherical features in the conversion layer. However, as we will demonstrate below, the impact of plasma density fluctuations on our calculations is negligible. 

It has been established that the spatial power spectrum of density fluctuations in the solar corona and solar wind follows the Kolmogorov power law~\cite{1989ApJ...337.1023C, 2001SSRv...97....9W, 2007ApJ...671..894T, 2008ApJ...676.1338T}:
\beq
\label{eq:fluctuation_P}
P(q) = C_N^2 q^{-\alpha}, q_o<q<q_i.
\eeq
Here, $\alpha = 11/3$ represents the slope of the Kolmogorov spectrum in three dimensions, where $q$ is the spatial wave number, and $l_{i,o} = 2\pi/q_{i,o}$ represent the inner and outer scales, respectively. The inner scale can be identified as the ion inertial scale~\cite{1989ApJ...337.1023C, 2008ApJ...676.1338T}.
In the solar corona, the outer scale $l_o$ is approximately $10^6$ times larger than $l_i$~\cite{2008ApJ...676.1338T}. Both $l_i$ and $l_o$ increase with distance $r$ from the Sun. By fitting the observational data summarized in\cite{1989ApJ...337.1023C, 2001SSRv...97....9W}, we can derive $l_i(r) \approx 0.1 (r/R_{\odot})^{1.8}$ km and $l_o(r) \approx 1.6 \times 10^5 (r/R_{\odot})^{0.8}$ km. Although the observation data used in Refs.~\cite{1989ApJ...337.1023C, 2001SSRv...97....9W} are within the range of approximately $1R_{\odot}$ to $100R_{\odot}$, the clear linear increase of $l_i$ and $l_o$ with distance in the logarithmic scale allows us to extend the applicable range of $l_i(r)$ and $l_o(r)$ to 1AU. Consequently, we obtain the ratio between the inner scale and the outer scale as follows:
\begin{equation}
\frac{q_i(r)}{q_o(r)} \approx 1.6 \times 10^6 \left(\frac{r}{R_{\odot}}\right)^{-1}.
\end{equation}
This ratio is approximately $10^6$ at $1R_{\odot}$, consistent with that used in Ref.~\cite{2008ApJ...676.1338T} for the solar corona, and decreases to about $7 \times 10^3$ at 1AU.

Integrating Eq.\eqref{eq:fluctuation_P} yields $\int_{q_o}^{q_i}d^3q P(q) = \left<\Delta n_e^2\right>$, where $\left<\Delta n_e^2\right>$ represents the variance of density fluctuations~\cite{2008ApJ...676.1338T}. From this integration, we can determine the normalization factor 
\begin{align}
C_N^2 \approx (\alpha-3)/(4\pi) \cdot q_o^{\alpha-3} \cdot \epsilon_e^2 n_e^2 .    
\end{align}
The dimensionless quantity $\epsilon_e = \Delta n_e/n_e$ denotes the relative density fluctuation. At large distances, $\epsilon_e$ remains approximately constant at about $0.07$ and is insensitive to radial distance variations~\cite{2007ApJ...671..894T, 2008ApJ...676.1338T, 2018ApJ...857...82K}.
However, at small distances within approximately $10R_{\odot}$, recent observations by the Parker Solar Probe indicate that $\epsilon_e \propto (r/R_{\odot})^{-0.55}$~\cite{2020ApJS..246...57K}. Consequently, we employ a piecewise function to characterize $\epsilon_e(r)$.

We perform a Fourier transformation of density fluctuations to obtain $\Delta n_e(r) = \int_{q_o}^{q_i} dq \Delta \tilde{n}_e(q) \mathrm{e}^{iqr}$. Consequently, we derive the average of radial derivatives as follows~\cite{An:2023wij},
\beq\label{eq:derivatives_average}
&\left<(n_e')^2\right>\simeq \left<(\Delta n_e')^2\right>
\simeq \frac{\alpha-3}{5-\alpha}\epsilon_e^2 n_e^2 q_o^{\alpha-3} q_i^{5-\alpha}, \\
&\left<(n_e'')^2\right>\simeq \left<(\Delta n_e'')^2\right>
\simeq \frac{\alpha-3}{7-\alpha}\epsilon_e^2 n_e^2 q_o^{\alpha-3} q_i^{7-\alpha}.
\eeq

\begin{figure}
    \centering
\includegraphics[width=1\linewidth]{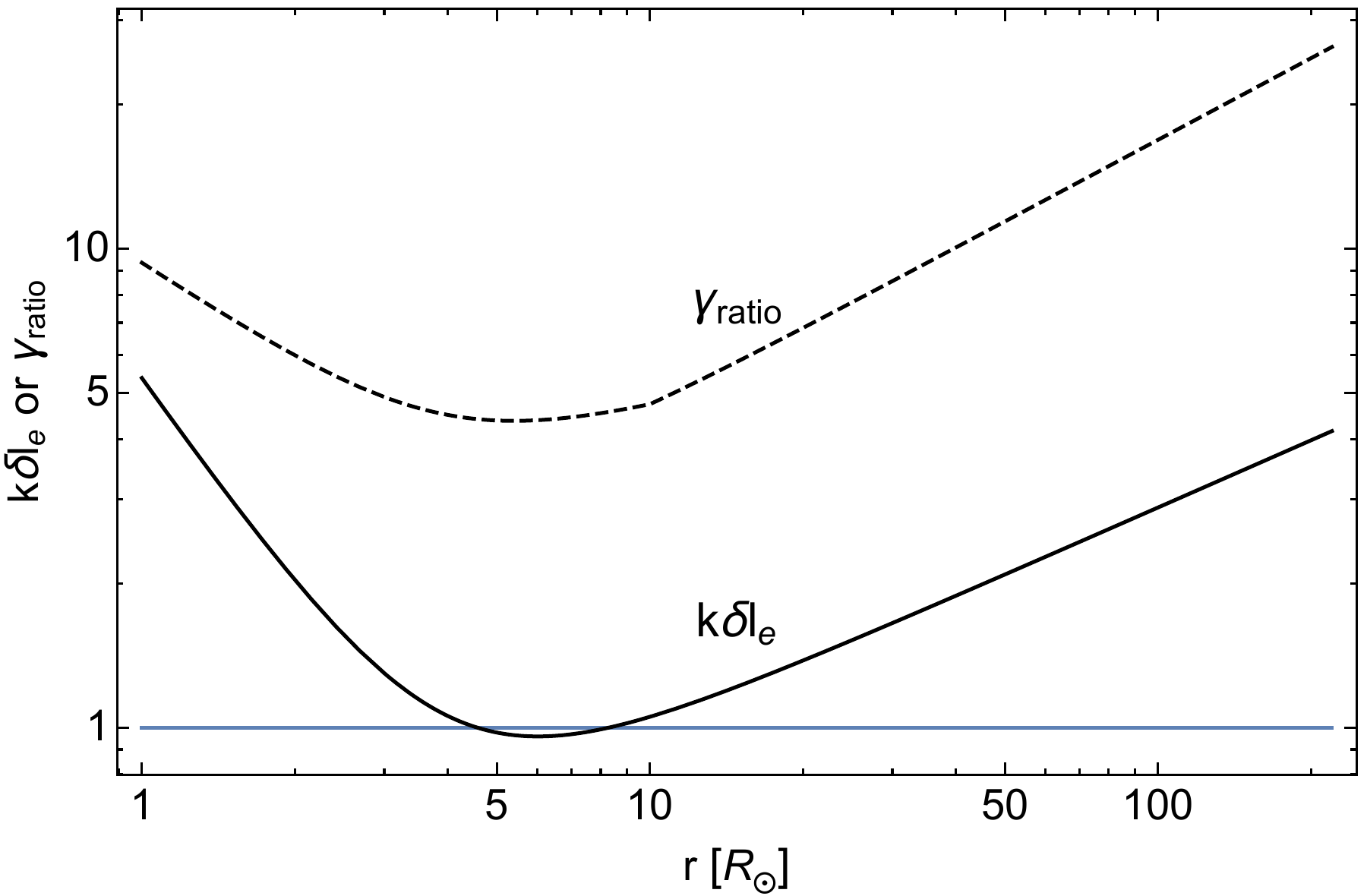}
    \caption{The effects of density fluctuations in the WKB approximation and the saddle-point method. The solid black line represents the value of $k \delta l_{e} $. The dashed black line represents the value of $\gamma_{\rm ratio}$. The blue line denotes the unity. 
    }
    \label{fig:effect_fluctuation}
\end{figure}

Firstly, we check whether the WKB approximation, applied in deriving Eq.~\eqref{eq:WKB_result}, remains valid. The typical variation length can be estimated as 
$\delta l_e \simeq |\epsilon_e n_e/n_e'|$. 
The WKB approximation is applicable if $\delta l_e$ is larger than $k^{-1}$ of dark photon.
Utilizing Eq.~\eqref{eq:derivatives_average}, we find
\beq
k \delta l_e  \simeq \left(\frac{\alpha-3}{5-\alpha}\right)^{-1/2}
\left(\frac{k}{q_i}\right)\left(\frac{q_i}{q_o}\right)^{1/3}. 
\eeq
Here, $k \simeq m_{A'} v_{\rm DM}$, and DPDM mass $m_{A'}$ equals the electron plasma frequency at the resonant layer, so $k(r)$ can be determined by Eq.~\eqref{eq:ne_profile}. The numerical result is depicted in Fig.~\ref{fig:effect_fluctuation}, 
demonstrating that the condition $k \delta l_e\gtrsim 1$ in the range from $1 ~R_{\odot}$ to $1$ AU is generally met.

Secondly, we assess whether the resonant length $\delta l_{\rm res}$ in Eq.~\eqref{eq:l_res} is smaller than $\delta l_e$. Here, $\delta l_{\rm res}$ represents the length of the region that contributes most to the conversion probability. If $\delta l_{\rm res}$ is smaller than $\delta l_e$, the resonant layer will not be significantly affected by the fluctuations. Eq.~\eqref{eq:l_res} can be expressed in terms of $\delta l_e$ as follows:
\beq
\delta l_{\rm res} \simeq \sqrt{\pi} v_{\rm DM} (\delta l_e/k)^{1/2},
\eeq
indicating that the ratio $\delta l_{\rm res}/\delta l_e \ll 1$ because $v_{\rm DM} \sim 10^{-3}c$ and $k \delta l_e \gtrsim 1$. Therefore, the resonant layer is minimally affected by the fluctuations.

Thirdly, when applying the saddle-point method, as expressed in Eq.~\eqref{eq:saddle-point-appro}, we retained the second derivative $f''$ while omitting higher derivative terms in the Taylor series. To demonstrate that the second derivative is indeed the dominant term, even in the presence of density fluctuations, we calculate the following quantity:
\begin{align}
\label{eq:third_to_two_ratio}
&\gamma_{\rm ratio} 
 \equiv  \frac{1/2! \cdot f''(r)}{1/3! \cdot f'''(r) \delta l_{\rm res}}
\\
&\simeq
\frac{3}{\sqrt{4\pi}}
\left(\frac{\alpha-3}{5-\alpha}\right)^{\frac{3}{4}}
\left(\frac{\alpha-3}{7-\alpha}\right)^{-\frac{1}{2}}
v_{\rm DM}^{-1}
\epsilon_e^{\frac{1}{2}}
\left(\frac{k}{q_i}\right)^{\frac{1}{2}}
\left(\frac{q_i}{q_o}\right)^{\frac{3-\alpha}{4}} . \nonumber
\end{align} 
The numeric value of $\gamma_{\rm ratio}$ is depicted in Fig.~\ref{fig:effect_fluctuation}, illustrating that the second derivative is indeed dominant. Additionally, the numerical results from Ref.~\cite{Brahma:2023zcw} indicate that the saddle-point method works with good accuracy for a ratio such as Eq.~\eqref{eq:third_to_two_ratio} larger than 1.
Therefore, the above arguments establish the validity of the expression for the conversion probability, Eq.\eqref{eq:prob}, in the presence of density fluctuations at a single resonant point. 

However, it is worth noting that density fluctuations may induce multiple resonant points that satisfy $\omega_{p} = m_{A'}$, potentially altering the magnitude of the flux of converted photons. Nevertheless, numeric simulations presented in Ref.~\cite{An:2023wij} demonstrate that the flux remains the same as in the case of no fluctuations.
This consistency can be understood as follows: the density fluctuations introduce two opposing effects. On one hand, additional resonant points may emerge, tending to increase the signal flux. On the other hand, the gradient of $n_e$ with respect to distance becomes steeper, potentially decreasing the signal flux. Ref.~\cite{An:2023wij} provides detailed numeric simulations illustrating that these two opposing effects effectively cancel each other out.
Lastly, as demonstrated in Ref.~\cite{An:2023wij}, the multiple resonant points concentrate within a narrow range, significantly smaller than the distance of the conversion layer to the solar center, $r_c$. Consequently, the non-spherical effects induced by density fluctuations are negligible.

In the last portion, we explore the possibility of two neighboring saddle points trapping the converted photons, preventing their propagation out and resulting in absorption by the plasma. The typical length scale of fluctuations is denoted by $\delta l_{\rm flu}$, while the resonant region is characterized by a length scale $\delta l_{\rm reg} \approx \epsilon_e n_e \cdot |dn_e/dr|^{-1}$. Therefore, the number of fluctuation steps can be calculated as $N_{\rm flu} = \delta l_{\rm reg}/ \delta l_{\rm flu}$. For the plasma density $n_e$, fluctuations randomly occur in both upward and downward directions.

Suppose we initiate the starting point with the resonant plasma density. The approximate number of times the fluctuation crosses the resonant density is $N_{\rm flu}^{1/2}$. Along the radial direction in one dimension, these points partition the trajectory into $N_{\rm flu}^{1/2}$ sections. Half of these sections represent dip regions, where $n_e < n_e^{\rm res}$, while the other half represent peak regions, where $n_e > n_e^{\rm res}$. On average, each section has a length of about $L_{\rm sec} = \delta l_{\rm region}/N_{\rm flu}^{1/2}$. As the resonant photon propagates, it can traverse the dip regions but encounters random scattering or reflection upon encountering the peak regions.

In three-dimensional random propagation, the photon eventually exits this region, albeit with an increased propagation time due to scattering. The random walk has a step size of $L_{\rm sec}$, and for each step, it takes the elapsed time $\delta t = L_{\rm sec}/ \bar{v}$, where $\bar{v}$ is the averaged speed of the photon when it crosses the dip region. Exactly at the resonant point, the speed of the converted photons is $v_{\rm DM}\sim 10^{-3}c$. However, when traversing the dip region, the speed quickly becomes relativistic because $n_e$ decreases to about $(1-\epsilon_e) n_e$. Therefore, with a relative fluctuation fraction of $\epsilon_e = 10\%$, we obtain an average speed of about $\bar{v} \sim 0.2 \,c$.
According to 3D random propagation, the relation between the trapping time and the variance of the final position is given as $t_{\rm trap} =  \sigma^2/L_{\rm sec}^2 \delta t$, where we should replace $\sigma$ with the length of the region $\delta l_{\rm region}$. After some algebra, we arrive at the trapping time being $t_{\rm trap} = \sqrt{N_{\rm flu}} \delta l_{\rm region}/ \bar{v}$.

Meanwhile, the absorption rate of converted photons due to the inverse bremsstrahlung process, $\Gamma_{\rm inv}$, is provided in Ref.~\cite{An:2020jmf}. Consequently, the probability of a converted photon being absorbed in the resonant region is characterized by $\Gamma \, t_{\rm trap}$. Numerical results of $\Gamma \, t_{\rm trap}$ are approximately of the order of one for the largest frequency $10^{-7}$~eV detectable by PSP. With decreasing frequency, this probability rapidly diminishes to insignificance. 
Moreover, it is important to note that these values apply to photons resonantly produced at the innermost resonant points. For photons produced at the outer region, the number of random-walking steps and the required variance $\sigma$ decrease, substantially shortening the trapping time. Therefore, we conclude that the trapping effect does not significantly alter our result.

In summary, the presence of plasma density fluctuations minimally affects our derivations of the conversion probability, the flux of converted photons, and the spherical shape of the conversion layer. Additionally, we estimated that the trapping effect on the converted photons is not significant. These findings are primarily attributed to the small magnitude of density fluctuations, characterized by $\epsilon_e \sim 10\%$, and to the fact that the fluctuations predominantly occur at larger scales, as dictated by the nature of the Kolmogorov spectrum.
\\

\bibliographystyle{utphys}
\bibliography{references}

\providecommand{\href}[2]{#2}\begingroup\raggedright\begin{thebibliography}{10}

\bibitem{Holdom:1985ag}
B.~Holdom, ``{Two U(1)'s and Epsilon Charge Shifts},''
\href{http://dx.doi.org/10.1016/0370-2693(86)91377-8}{{\em Phys. Lett.}
  {\bfseries 166B} (1986) 196--198}.

\bibitem{Dienes:1996zr}
K.~R. Dienes, C.~F. Kolda, and J.~March-Russell, ``{Kinetic mixing and the
  supersymmetric gauge hierarchy},''
  \href{http://dx.doi.org/10.1016/S0550-3213(97)00173-9}{{\em Nucl. Phys. B}
  {\bfseries 492} (1997) 104--118},
  \href{http://arxiv.org/abs/hep-ph/9610479}{{\ttfamily arXiv:hep-ph/9610479}}.

\bibitem{Abel:2003ue}
S.~A. Abel and B.~W. Schofield, ``{Brane anti-brane kinetic mixing,
  millicharged particles and SUSY breaking},''
  \href{http://dx.doi.org/10.1016/j.nuclphysb.2004.02.037}{{\em Nucl. Phys. B}
  {\bfseries 685} (2004) 150--170},
  \href{http://arxiv.org/abs/hep-th/0311051}{{\ttfamily arXiv:hep-th/0311051}}.

\bibitem{Abel:2006qt}
S.~A. Abel, J.~Jaeckel, V.~V. Khoze, and A.~Ringwald, ``{Illuminating the
  Hidden Sector of String Theory by Shining Light through a Magnetic Field},''
  \href{http://dx.doi.org/10.1016/j.physletb.2008.03.076}{{\em Phys. Lett. B}
  {\bfseries 666} (2008) 66--70},
  \href{http://arxiv.org/abs/hep-ph/0608248}{{\ttfamily arXiv:hep-ph/0608248}}.

\bibitem{Nelson:2011sf}
A.~E. Nelson and J.~Scholtz, ``{Dark Light, Dark Matter and the Misalignment
  Mechanism},'' \href{http://dx.doi.org/10.1103/PhysRevD.84.103501}{{\em Phys.
  Rev.} {\bfseries D84} (2011) 103501},
\href{http://arxiv.org/abs/1105.2812}{{\ttfamily arXiv:1105.2812 [hep-ph]}}.

\bibitem{Arias:2012az}
P.~Arias, D.~Cadamuro, M.~Goodsell, J.~Jaeckel, J.~Redondo, and A.~Ringwald,
  ``{WISPy Cold Dark Matter},''
  \href{http://dx.doi.org/10.1088/1475-7516/2012/06/013}{{\em JCAP} {\bfseries
  1206} (2012) 013},
\href{http://arxiv.org/abs/1201.5902}{{\ttfamily arXiv:1201.5902 [hep-ph]}}.

\bibitem{AlonsoAlvarez:2019cgw}
G.~Alonso-Álvarez, T.~Hugle, and J.~Jaeckel, ``{Misalignment \& Co.:
  (Pseudo-)scalar and vector dark matter with curvature couplings},''
\href{http://arxiv.org/abs/1905.09836}{{\ttfamily arXiv:1905.09836 [hep-ph]}}.

\bibitem{Nakayama:2019rhg}
K.~Nakayama, ``{Vector Coherent Oscillation Dark Matter},''
  \href{http://dx.doi.org/10.1088/1475-7516/2019/10/019}{{\em JCAP} {\bfseries
  1910} (2019) 019},
\href{http://arxiv.org/abs/1907.06243}{{\ttfamily arXiv:1907.06243 [hep-ph]}}.

\bibitem{Nakayama:2020rka}
K.~Nakayama, ``{Constraint on Vector Coherent Oscillation Dark Matter with
  Kinetic Function},''
  \href{http://dx.doi.org/10.1088/1475-7516/2020/08/033}{{\em JCAP} {\bfseries
  08} (2020) 033}, \href{http://arxiv.org/abs/2004.10036}{{\ttfamily
  arXiv:2004.10036 [hep-ph]}}.

\bibitem{Graham:2015rva}
P.~W. Graham, J.~Mardon, and S.~Rajendran, ``{Vector Dark Matter from
  Inflationary Fluctuations},''
  \href{http://dx.doi.org/10.1103/PhysRevD.93.103520}{{\em Phys. Rev.}
  {\bfseries D93} no.~10, (2016) 103520},
\href{http://arxiv.org/abs/1504.02102}{{\ttfamily arXiv:1504.02102 [hep-ph]}}.

\bibitem{Ema:2019yrd}
Y.~Ema, K.~Nakayama, and Y.~Tang, ``{Production of Purely Gravitational Dark
  Matter: The Case of Fermion and Vector Boson},''
  \href{http://dx.doi.org/10.1007/JHEP07(2019)060}{{\em JHEP} {\bfseries 07}
  (2019) 060}, \href{http://arxiv.org/abs/1903.10973}{{\ttfamily
  arXiv:1903.10973 [hep-ph]}}.

\bibitem{Kolb:2020fwh}
E.~W. Kolb and A.~J. Long, ``{Completely dark photons from gravitational
  particle production during the inflationary era},''
  \href{http://dx.doi.org/10.1007/JHEP03(2021)283}{{\em JHEP} {\bfseries 03}
  (2021) 283}, \href{http://arxiv.org/abs/2009.03828}{{\ttfamily
  arXiv:2009.03828 [astro-ph.CO]}}.

\bibitem{Salehian:2020asa}
B.~Salehian, M.~A. Gorji, H.~Firouzjahi, and S.~Mukohyama, ``{Vector dark
  matter production from inflation with symmetry breaking},''
  \href{http://dx.doi.org/10.1103/PhysRevD.103.063526}{{\em Phys. Rev. D}
  {\bfseries 103} no.~6, (2021) 063526},
  \href{http://arxiv.org/abs/2010.04491}{{\ttfamily arXiv:2010.04491
  [hep-ph]}}.

\bibitem{Ahmed:2020fhc}
A.~Ahmed, B.~Grzadkowski, and A.~Socha, ``{Gravitational production of vector
  dark matter},'' \href{http://dx.doi.org/10.1007/JHEP08(2020)059}{{\em JHEP}
  {\bfseries 08} (2020) 059}, \href{http://arxiv.org/abs/2005.01766}{{\ttfamily
  arXiv:2005.01766 [hep-ph]}}.

\bibitem{Nakai:2020cfw}
Y.~Nakai, R.~Namba, and Z.~Wang, ``{Light Dark Photon Dark Matter from
  Inflation},'' \href{http://dx.doi.org/10.1007/JHEP12(2020)170}{{\em JHEP}
  {\bfseries 12} (2020) 170}, \href{http://arxiv.org/abs/2004.10743}{{\ttfamily
  arXiv:2004.10743 [hep-ph]}}.

\bibitem{Nakayama:2020ikz}
K.~Nakayama and Y.~Tang, ``{Gravitational Production of Hidden Photon Dark
  Matter in Light of the XENON1T Excess},''
  \href{http://dx.doi.org/10.1016/j.physletb.2020.135977}{{\em Phys. Lett. B}
  {\bfseries 811} (2020) 135977},
  \href{http://arxiv.org/abs/2006.13159}{{\ttfamily arXiv:2006.13159
  [hep-ph]}}.

\bibitem{Firouzjahi:2020whk}
H.~Firouzjahi, M.~A. Gorji, S.~Mukohyama, and B.~Salehian, ``{Dark photon dark
  matter from charged inflaton},''
  \href{http://dx.doi.org/10.1007/JHEP06(2021)050}{{\em JHEP} {\bfseries 06}
  (2021) 050}, \href{http://arxiv.org/abs/2011.06324}{{\ttfamily
  arXiv:2011.06324 [hep-ph]}}.

\bibitem{Bastero-Gil:2021wsf}
M.~Bastero-Gil, J.~Santiago, L.~Ubaldi, and R.~Vega-Morales, ``{Dark photon
  dark matter from a rolling inflaton},''
  \href{http://dx.doi.org/10.1088/1475-7516/2022/02/015}{{\em JCAP} {\bfseries
  02} no.~02, (2022) 015}, \href{http://arxiv.org/abs/2103.12145}{{\ttfamily
  arXiv:2103.12145 [hep-ph]}}.

\bibitem{Firouzjahi:2021lov}
H.~Firouzjahi, M.~A. Gorji, S.~Mukohyama, and A.~Talebian, ``{Dark matter from
  entropy perturbations in curved field space},''
  \href{http://dx.doi.org/10.1103/PhysRevD.105.043501}{{\em Phys. Rev. D}
  {\bfseries 105} no.~4, (2022) 043501},
  \href{http://arxiv.org/abs/2110.09538}{{\ttfamily arXiv:2110.09538 [gr-qc]}}.

\bibitem{Sato:2022jya}
T.~Sato, F.~Takahashi, and M.~Yamada, ``{Gravitational production of dark
  photon dark matter with mass generated by the Higgs mechanism},''
  \href{http://arxiv.org/abs/2204.11896}{{\ttfamily arXiv:2204.11896
  [hep-ph]}}.

\bibitem{Co:2018lka}
R.~T. Co, A.~Pierce, Z.~Zhang, and Y.~Zhao, ``{Dark Photon Dark Matter Produced
  by Axion Oscillations},''
\href{http://arxiv.org/abs/1810.07196}{{\ttfamily arXiv:1810.07196 [hep-ph]}}.

\bibitem{Dror:2018pdh}
J.~A. Dror, K.~Harigaya, and V.~Narayan, ``{Parametric Resonance Production of
  Ultralight Vector Dark Matter},''
\href{http://arxiv.org/abs/1810.07195}{{\ttfamily arXiv:1810.07195 [hep-ph]}}.

\bibitem{Bastero-Gil:2018uel}
M.~Bastero-Gil, J.~Santiago, L.~Ubaldi, and R.~Vega-Morales, ``{Vector dark
  matter production at the end of inflation},''
\href{http://arxiv.org/abs/1810.07208}{{\ttfamily arXiv:1810.07208 [hep-ph]}}.

\bibitem{Agrawal:2018vin}
P.~Agrawal, N.~Kitajima, M.~Reece, T.~Sekiguchi, and F.~Takahashi, ``{Relic
  Abundance of Dark Photon Dark Matter},''
\href{http://arxiv.org/abs/1810.07188}{{\ttfamily arXiv:1810.07188 [hep-ph]}}.

\bibitem{Co:2021rhi}
R.~T. Co, K.~Harigaya, and A.~Pierce, ``{Gravitational waves and dark photon
  dark matter from axion rotations},''
  \href{http://dx.doi.org/10.1007/JHEP12(2021)099}{{\em JHEP} {\bfseries 12}
  (2021) 099}, \href{http://arxiv.org/abs/2104.02077}{{\ttfamily
  arXiv:2104.02077 [hep-ph]}}.

\bibitem{Nakayama:2021avl}
K.~Nakayama and W.~Yin, ``{Hidden photon and axion dark matter from symmetry
  breaking},'' \href{http://dx.doi.org/10.1007/JHEP10(2021)026}{{\em JHEP}
  {\bfseries 10} (2021) 026}, \href{http://arxiv.org/abs/2105.14549}{{\ttfamily
  arXiv:2105.14549 [hep-ph]}}.

\bibitem{Cyncynates:2023zwj}
D.~Cyncynates and Z.~J. Weiner, ``{Detectable, defect-free dark photon dark
  matter},'' \href{http://arxiv.org/abs/2310.18397}{{\ttfamily arXiv:2310.18397
  [hep-ph]}}.

\bibitem{Long:2019lwl}
A.~J. Long and L.-T. Wang, ``{Dark Photon Dark Matter from a Network of Cosmic
  Strings},''
\href{http://arxiv.org/abs/1901.03312}{{\ttfamily arXiv:1901.03312 [hep-ph]}}.

\bibitem{An:2020jmf}
H.~An, F.~P. Huang, J.~Liu, and W.~Xue, ``{Radio-frequency Dark Photon Dark
  Matter across the Sun},''
  \href{http://dx.doi.org/10.1103/PhysRevLett.126.181102}{{\em Phys. Rev.
  Lett.} {\bfseries 126} no.~18, (2021) 181102},
  \href{http://arxiv.org/abs/2010.15836}{{\ttfamily arXiv:2010.15836
  [hep-ph]}}.

\bibitem{An:2023wij}
H.~An, X.~Chen, S.~Ge, J.~Liu, and Y.~Luo, ``{Searching for ultralight dark
  matter conversion in solar corona using Low Frequency Array data},''
  \href{http://dx.doi.org/10.1038/s41467-024-45033-4}{{\em Nature Commun.}
  {\bfseries 15} no.~1, (2024) 915},
  \href{http://arxiv.org/abs/2301.03622}{{\ttfamily arXiv:2301.03622
  [hep-ph]}}.

\bibitem{2016Fox}
N.~J. {Fox}, M.~C. {Velli}, S.~D. {Bale}, R.~{Decker}, A.~{Driesman}, R.~A.
  {Howard}, J.~C. {Kasper}, J.~{Kinnison}, M.~{Kusterer}, D.~{Lario}, M.~K.
  {Lockwood}, D.~J. {McComas}, N.~E. {Raouafi}, and A.~{Szabo}, ``{The Solar
  Probe Plus Mission: Humanity's First Visit to Our Star},''
  \href{http://dx.doi.org/10.1007/s11214-015-0211-6}{{\em Space Science
  Reviews} {\bfseries 204} no.~1-4, (Dec., 2016) 7--48}.

\bibitem{PSPorbitInfo}
 \url{http://parkersolarprobe.jhuapl.edu/The-Mission/index.php#Where-Is-PSP}.

\bibitem{2008Kaiser_SSRv}
M.~L. {Kaiser}, T.~A. {Kucera}, J.~M. {Davila}, O.~C. {St. Cyr},
  M.~{Guhathakurta}, and E.~{Christian}, ``{The STEREO Mission: An
  Introduction},'' \href{http://dx.doi.org/10.1007/s11214-007-9277-0}{{\em
  Space Science Reviews} {\bfseries 136} no.~1-4, (Apr., 2008) 5--16}.

\bibitem{Witte:2020rvb}
S.~J. Witte, S.~Rosauro-Alcaraz, S.~D. McDermott, and V.~Poulin, ``{Dark photon
  dark matter in the presence of inhomogeneous structure},''
  \href{http://dx.doi.org/10.1007/JHEP06(2020)132}{{\em JHEP} {\bfseries 06}
  (2020) 132}, \href{http://arxiv.org/abs/2003.13698}{{\ttfamily
  arXiv:2003.13698 [astro-ph.CO]}}.

\bibitem{Raffelt:1987im}
G.~Raffelt and L.~Stodolsky, ``{Mixing of the Photon with Low Mass
  Particles},''
\href{http://dx.doi.org/10.1103/PhysRevD.37.1237}{{\em Phys. Rev.} {\bfseries
  D37} (1988) 1237}.

\bibitem{Supp_cite}
 See Supplemental Material, which includes Refs. [37-75], for more information
  on the technical details, the satellites, and data analysis.

\bibitem{Witte:2021arp}
S.~J. Witte, D.~Noordhuis, T.~D.~P. Edwards, and C.~Weniger, ``{Axion-photon
  conversion in neutron star magnetospheres: The role of the plasma in the
  Goldreich-Julian model},''
  \href{http://dx.doi.org/10.1103/PhysRevD.104.103030}{{\em Phys. Rev. D}
  {\bfseries 104} no.~10, (2021) 103030},
  \href{http://arxiv.org/abs/2104.07670}{{\ttfamily arXiv:2104.07670
  [hep-ph]}}.

\bibitem{Mirizzi:2009iz}
A.~Mirizzi, J.~Redondo, and G.~Sigl, ``{Microwave Background Constraints on
  Mixing of Photons with Hidden Photons},''
  \href{http://dx.doi.org/10.1088/1475-7516/2009/03/026}{{\em JCAP} {\bfseries
  03} (2009) 026}, \href{http://arxiv.org/abs/0901.0014}{{\ttfamily
  arXiv:0901.0014 [hep-ph]}}.

\bibitem{Drukier:1986tm}
A.~K. Drukier, K.~Freese, and D.~N. Spergel, ``{Detecting Cold Dark Matter
  Candidates},'' \href{http://dx.doi.org/10.1103/PhysRevD.33.3495}{{\em Phys.
  Rev. D} {\bfseries 33} (1986) 3495--3508}.

\bibitem{Choi:2013eda}
K.~Choi, C.~Rott, and Y.~Itow, ``{Impact of the dark matter velocity
  distribution on capture rates in the Sun},''
  \href{http://dx.doi.org/10.1088/1475-7516/2014/05/049}{{\em JCAP} {\bfseries
  05} (2014) 049}, \href{http://arxiv.org/abs/1312.0273}{{\ttfamily
  arXiv:1312.0273 [astro-ph.HE]}}.

\bibitem{Evans:2018bqy}
N.~W. Evans, C.~A.~J. O'Hare, and C.~McCabe, ``{Refinement of the standard halo
  model for dark matter searches in light of the Gaia Sausage},''
  \href{http://dx.doi.org/10.1103/PhysRevD.99.023012}{{\em Phys. Rev. D}
  {\bfseries 99} no.~2, (2019) 023012},
  \href{http://arxiv.org/abs/1810.11468}{{\ttfamily arXiv:1810.11468
  [astro-ph.GA]}}.

\bibitem{Hardy:2022ufh}
E.~Hardy and N.~Song, ``{Listening for dark photon radio signals from the
  Galactic Center},'' \href{http://dx.doi.org/10.1103/PhysRevD.107.115035}{{\em
  Phys. Rev. D} {\bfseries 107} no.~11, (2023) 115035},
  \href{http://arxiv.org/abs/2212.09756}{{\ttfamily arXiv:2212.09756
  [hep-ph]}}.

\bibitem{favorite2016solid}
J.~A. Favorite, ``The solid angle (geometry factor) for a spherical surface
  source and an arbitrary detector aperture,'' {\em Nuclear Instruments and
  Methods in Physics Research Section A: Accelerators, Spectrometers, Detectors
  and Associated Equipment} {\bfseries 813} (2016) 29--35.

\bibitem{manning2000instrumentation}
R.~Manning, ``Instrumentation for space-based low frequency radio astronomy,''
  {\em Geophysical monograph} {\bfseries 119} (2000) 329--337.

\bibitem{zarka2004jupiter}
P.~Zarka, B.~Cecconi, and W.~S. Kurth, ``Jupiter's low-frequency radio spectrum
  from cassini/radio and plasma wave science (rpws) absolute flux density
  measurements,'' {\em Journal of Geophysical Research: Space Physics}
  {\bfseries 109} no.~A9, (2004) .

\bibitem{eastwood2009measurements}
J.~P. Eastwood, S.~Bale, M.~Maksimovic, I.~Zouganelis, K.~Goetz, M.~Kaiser, and
  J.-L. Bougeret, ``Measurements of stray antenna capacitance in the
  stereo/waves instrument: Comparison of the radio frequency voltage spectrum
  with models of the galactic nonthermal continuum spectrum,'' {\em Radio
  Science} {\bfseries 44} no.~04, (2009) 1--8.

\bibitem{PSPdataLevel3}
 \url{https://research.ssl.berkeley.edu/data/psp/data/sci/fields/l3/}.

\bibitem{Moncuquet2020First/FIELDS}
M.~Moncuquet, N.~Meyer-Vernet, K.~Issautier, M.~Pulupa, J.~W. Bonnell, S.~D.
  Bale, T.~D. de~Wit, K.~Goetz, L.~Griton, P.~R. Harvey, R.~J. MacDowall,
  M.~Maksimovic, and D.~M. Malaspina, ``{First In Situ Measurements of Electron
  Density and Temperature from Quasi-thermal Noise Spectroscopy with Parker
  Solar Probe /FIELDS},''
  \href{http://dx.doi.org/10.3847/1538-4365/ab5a84}{{\em The Astrophysical
  Journal Supplement Series} {\bfseries 246} no.~2, (2, 2020) 44}.
  \url{https://iopscience.iop.org/article/10.3847/1538-4365/ab5a84
  https://iopscience.iop.org/article/10.3847/1538-4365/ab5a84/meta}.

\bibitem{balanis2015antenna}
C.~A. Balanis, {\em Antenna theory: analysis and design}.
\newblock John wiley \& sons, 2015.

\bibitem{Maksimovic2020AnticorrelationHelios}
M.~{Maksimovic}, S.~D. {Bale}, L.~{Ber{\v{c}}i{\v{c}}}, J.~W. {Bonnell}, A.~W.
  {Case}, T.~{Dudok de Wit}, K.~{Goetz}, J.~S. {Halekas}, P.~R. {Harvey},
  K.~{Issautier}, J.~C. {Kasper}, K.~E. {Korreck}, V.~K. {Jagarlamudi},
  N.~{Lahmiti}, D.~E. {Larson}, A.~{Lecacheux}, R.~{Livi}, R.~J. {MacDowall},
  D.~M. {Malaspina}, M.~M. {Martinovi{\'c}}, N.~{Meyer-Vernet}, M.~{Moncuquet},
  M.~{Pulupa}, C.~{Salem}, M.~L. {Stevens}, {\v{S}}.~{{\v{S}}tver{\'a}k},
  M.~{Velli}, and P.~L. {Whittlesey}, ``{Anticorrelation between the Bulk Speed
  and the Electron Temperature in the Pristine Solar Wind: First Results from
  the Parker Solar Probe and Comparison with Helios},''
  \href{http://dx.doi.org/10.3847/1538-4365/ab61fc}{{\em The Astrophysical
  Journal Supplement Series} {\bfseries 246} no.~2, (Feb., 2020) 62}.

\bibitem{SQTN_WEB}
 \url{https://cdpp-archive.cnes.fr/}.

\bibitem{PSP_FIELDS_Release_Note}
 \url{https://fields.ssl.berkeley.edu/}.

\bibitem{2017meyervernet}
N.~Meyer-Vernet, K.~Issautier, and M.~Moncuquet, ``{Quasi-thermal noise
  spectroscopy: The art and the practice},''
  \href{http://dx.doi.org/10.1002/2017JA024449}{{\em Journal of Geophysical
  Research: Space Physics} {\bfseries 122} no.~8, (2017) 7925--7945}.
  \url{https://hal.sorbonne-universite.fr/hal-01628354}.

\bibitem{Kasper2016}
J.~C. {Kasper}, R.~{Abiad}, G.~{Austin}, M.~{Balat-Pichelin}, S.~D. {Bale},
  J.~W. {Belcher}, P.~{Berg}, H.~{Bergner}, M.~{Berthomier}, J.~{Bookbinder},
  E.~{Brodu}, D.~{Caldwell}, A.~W. {Case}, B.~D.~G. {Chand ran}, P.~{Cheimets},
  J.~W. {Cirtain}, S.~R. {Cranmer}, D.~W. {Curtis}, P.~{Daigneau}, G.~{Dalton},
  B.~{Dasgupta}, D.~{DeTomaso}, M.~{Diaz-Aguado}, B.~{Djordjevic},
  B.~{Donaskowski}, M.~{Effinger}, V.~{Florinski}, N.~{Fox}, M.~{Freeman},
  D.~{Gallagher}, S.~P. {Gary}, T.~{Gauron}, R.~{Gates}, M.~{Goldstein},
  L.~{Golub}, D.~A. {Gordon}, R.~{Gurnee}, G.~{Guth}, J.~{Halekas}, K.~{Hatch},
  J.~{Heerikuisen}, G.~{Ho}, Q.~{Hu}, G.~{Johnson}, S.~P. {Jordan}, K.~E.
  {Korreck}, D.~{Larson}, A.~J. {Lazarus}, G.~{Li}, R.~{Livi}, M.~{Ludlam},
  M.~{Maksimovic}, J.~P. {McFadden}, W.~{Marchant}, B.~A. {Maruca}, D.~J.
  {McComas}, L.~{Messina}, T.~{Mercer}, S.~{Park}, A.~M. {Peddie},
  N.~{Pogorelov}, M.~J. {Reinhart}, J.~D. {Richardson}, M.~{Robinson},
  I.~{Rosen}, R.~M. {Skoug}, A.~{Slagle}, J.~T. {Steinberg}, M.~L. {Stevens},
  A.~{Szabo}, E.~R. {Taylor}, C.~{Tiu}, P.~{Turin}, M.~{Velli}, G.~{Webb},
  P.~{Whittlesey}, K.~{Wright}, S.~T. {Wu}, and G.~{Zank}, ``{Solar Wind
  Electrons Alphas and Protons (SWEAP) Investigation: Design of the Solar Wind
  and Coronal Plasma Instrument Suite for Solar Probe Plus},''
  \href{http://dx.doi.org/10.1007/s11214-015-0206-3}{{\em Space Science
  Reviews} {\bfseries 204} no.~1-4, (Dec., 2016) 131--186}.

\bibitem{Maksimovic2020}
M.~{Maksimovic}, S.~D. {Bale}, L.~{Ber{\v{c}}i{\v{c}}}, J.~W. {Bonnell}, A.~W.
  {Case}, T.~D.~d. {Wit}, K.~{Goetz}, J.~S. {Halekas}, P.~R. {Harvey},
  K.~{Issautier}, J.~C. {Kasper}, K.~E. {Korreck}, V.~K. {Jagarlamudi},
  N.~{Lahmiti}, D.~E. {Larson}, A.~{Lecacheux}, R.~{Livi}, R.~J. {MacDowall},
  D.~M. {Malaspina}, M.~M. {Martinovi{\'c}}, N.~{Meyer-Vernet}, M.~{Moncuquet},
  M.~{Pulupa}, C.~{Salem}, M.~L. {Stevens}, {\v{S}}.~{{\v{S}}tver{\'a}k},
  M.~{Velli}, and P.~L. {Whittlesey}, ``{Anticorrelation between the Bulk Speed
  and the Electron Temperature in the Pristine Solar Wind: First Results from
  the Parker Solar Probe and Comparison with Helios},''
  \href{http://dx.doi.org/10.3847/1538-4365/ab61fc}{{\em The Astrophysical
  Journal Supplement Series} {\bfseries 246} no.~2, (Feb., 2020) 62}.

\bibitem{2022Martinovic}
M.~M. {Martinovi{\'c}}, A.~R. {Dordevi{\'c}}, K.~G. {Klein},
  M.~{Maksimovi{\'c}}, K.~{Issautier}, M.~{Liu}, M.~{Pulupa}, S.~D. {Bale},
  J.~S. {Halekas}, and M.~D. {McManus}, ``{Plasma Parameters From Quasi-Thermal
  Noise Observed by Parker Solar Probe: A New Model for the Antenna
  Response},'' \href{http://dx.doi.org/10.1029/2021JA030182}{{\em Journal of
  Geophysical Research (Space Physics)} {\bfseries 127} no.~4, (Apr., 2022)
  e30182}.

\bibitem{2023Liu_QTN}
M.~{Liu}, K.~{Issautier}, M.~{Moncuquet}, N.~{Meyer-Vernet}, M.~{Maksimovic},
  J.~{Huang}, M.~M. {Martinovic}, L.~{Griton}, N.~{Chrysaphi}, V.~K.
  {Jagarlamudi}, S.~D. {Bale}, M.~{Pulupa}, J.~C. {Kasper}, and M.~L.
  {Stevens}, ``{Total electron temperature derived from quasi-thermal noise
  spectroscopy in the pristine solar wind from Parker Solar Probe
  observations},'' \href{http://dx.doi.org/10.1051/0004-6361/202245450}{{\em
  Astronomy \& Astrophysics} {\bfseries 674} (June, 2023) A49},
  \href{http://arxiv.org/abs/2303.11035}{{\ttfamily arXiv:2303.11035
  [astro-ph.SR]}}.

\bibitem{Zheng_2024}
X.~Zheng, K.~Liu, M.~M. Martinović, V.~Pierrard, M.~Liu, Q.~He, K.~Cheng,
  Y.~Liu, and Y.~Wang, ``Solar wind density and core temperature derived from
  the psp quasi-thermal noise measurements,''
  \href{http://dx.doi.org/10.3847/1538-4357/ad236d}{{\em The Astrophysical
  Journal} {\bfseries 963} no.~2, (Mar, 2024) 154}.
  \url{https://dx.doi.org/10.3847/1538-4357/ad236d}.

\bibitem{Case2020}
A.~W. {Case}, J.~C. {Kasper}, M.~L. {Stevens}, K.~E. {Korreck}, K.~{Paulson},
  P.~{Daigneau}, D.~{Caldwell}, M.~{Freeman}, T.~{Henry}, B.~{Klingensmith},
  J.~A. {Bookbinder}, M.~{Robinson}, P.~{Berg}, C.~{Tiu}, J.~{Wright}, K.~H.,
  M.~J. {Reinhart}, D.~{Curtis}, M.~{Ludlam}, D.~{Larson}, P.~{Whittlesey},
  R.~{Livi}, K.~G. {Klein}, and M.~M. {Martinovi{\'c}}, ``{The Solar Probe Cup
  on the Parker Solar Probe},''
  \href{http://dx.doi.org/10.3847/1538-4365/ab5a7b}{{\em The Astrophysical
  Journal Supplement Series} {\bfseries 246} no.~2, (Feb., 2020) 43},
  \href{http://arxiv.org/abs/1912.02581}{{\ttfamily arXiv:1912.02581
  [astro-ph.IM]}}.

\bibitem{2022Livi}
R.~{Livi}, D.~E. {Larson}, J.~C. {Kasper}, R.~{Abiad}, A.~W. {Case}, K.~G.
  {Klein}, D.~W. {Curtis}, G.~{Dalton}, M.~{Stevens}, K.~E. {Korreck}, G.~{Ho},
  M.~{Robinson}, C.~{Tiu}, P.~L. {Whittlesey}, J.~L. {Verniero}, J.~{Halekas},
  J.~{McFadden}, M.~{Marckwordt}, A.~{Slagle}, M.~{Abatcha}, A.~{Rahmati}, and
  M.~D. {McManus}, ``{The Solar Probe ANalyzer-Ions on the Parker Solar
  Probe},'' \href{http://dx.doi.org/10.3847/1538-4357/ac93f5}{{\em \apj}
  {\bfseries 938} no.~2, (Oct., 2022) 138}.

\bibitem{Whittlesey_2020}
P.~L. Whittlesey, D.~E. Larson, J.~C. Kasper, J.~Halekas, M.~Abatcha, R.~Abiad,
  M.~Berthomier, A.~W. Case, J.~Chen, D.~W. Curtis, G.~Dalton, K.~G. Klein,
  K.~E. Korreck, R.~Livi, M.~Ludlam, M.~Marckwordt, A.~Rahmati, M.~Robinson,
  A.~Slagle, M.~L. Stevens, C.~Tiu, and J.~L. Verniero, ``The solar probe
  {ANalyzers}{\textemdash}electrons on the parker solar probe,''
  \href{http://dx.doi.org/10.3847/1538-4365/ab7370}{{\em The Astrophysical
  Journal Supplement Series} {\bfseries 246} no.~2, (Mar, 2020) 74}.
  \url{https://doi.org/10.3847%2F1538-4365%2Fab7370}.

\bibitem{2021Kasper}
J.~C. {Kasper}, K.~G. {Klein}, E.~{Lichko}, J.~{Huang}, C.~H.~K. {Chen}, S.~T.
  {Badman}, J.~{Bonnell}, P.~L. {Whittlesey}, R.~{Livi}, D.~{Larson},
  M.~{Pulupa}, A.~{Rahmati}, D.~{Stansby}, K.~E. {Korreck}, M.~{Stevens}, A.~W.
  {Case}, S.~D. {Bale}, M.~{Maksimovic}, M.~{Moncuquet}, K.~{Goetz}, J.~S.
  {Halekas}, D.~{Malaspina}, N.~E. {Raouafi}, A.~{Szabo}, R.~{MacDowall},
  M.~{Velli}, T.~{Dudok de Wit}, and G.~P. {Zank}, ``{Parker Solar Probe Enters
  the Magnetically Dominated Solar Corona},''
  \href{http://dx.doi.org/10.1103/PhysRevLett.127.255101}{{\em \prl} {\bfseries
  127} no.~25, (Dec., 2021) 255101}.

\bibitem{2021Zhao}
S.~Q. {Zhao}, H.~{Yan}, T.~Z. {Liu}, M.~{Liu}, and M.~{Shi}, ``{Analysis of
  Magnetohydrodynamic Perturbations in the Radial-field Solar Wind from Parker
  Solar Probe Observations},''
  \href{http://dx.doi.org/10.3847/1538-4357/ac2ffe}{{\em \apj} {\bfseries 923}
  no.~2, (Dec., 2021) 253}, \href{http://arxiv.org/abs/2106.03807}{{\ttfamily
  arXiv:2106.03807 [astro-ph.SR]}}.

\bibitem{2021Liu}
M.~{Liu}, K.~{Issautier}, N.~{Meyer-Vernet}, M.~{Moncuquet}, M.~{Maksimovic},
  J.~S. {Halekas}, J.~{Huang}, L.~{Griton}, S.~{Bale}, J.~W. {Bonnell}, A.~W.
  {Case}, K.~{Goetz}, P.~R. {Harvey}, J.~C. {Kasper}, R.~J. {MacDowall}, D.~M.
  {Malaspina}, M.~{Pulupa}, and M.~L. {Stevens}, ``{Solar wind energy flux
  observations in the inner heliosphere: first results from Parker Solar
  Probe},'' \href{http://dx.doi.org/10.1051/0004-6361/202039615}{{\em Astronomy
  \& Astrophysics} {\bfseries 650} (June, 2021) A14},
  \href{http://arxiv.org/abs/2101.03121}{{\ttfamily arXiv:2101.03121
  [astro-ph.SR]}}.

\bibitem{2021ApJLiu}
Y.~D. {Liu}, C.~{Chen}, M.~L. {Stevens}, and M.~{Liu}, ``{Determination of
  Solar Wind Angular Momentum and Alfv{\'e}n Radius from Parker Solar Probe
  Observations},'' \href{http://dx.doi.org/10.3847/2041-8213/abe38e}{{\em The
  Astrophysical Journal Letters} {\bfseries 908} no.~2, (Feb., 2021) L41},
  \href{http://arxiv.org/abs/2102.03376}{{\ttfamily arXiv:2102.03376
  [astro-ph.SR]}}.

\bibitem{2024LiuYingD}
Y.~D. {Liu}, B.~{Zhu}, H.~{Ran}, H.~{Hu}, M.~{Liu}, X.~{Zhao}, R.~{Wang}, M.~L.
  {Stevens}, and S.~D. {Bale}, ``{Direct In Situ Measurements of a Fast Coronal
  Mass Ejection and Associated Structures in the Corona},''
  \href{http://dx.doi.org/10.48550/arXiv.2401.06449}{{\em arXiv e-prints}
  (Jan., 2024) arXiv:2401.06449},
  \href{http://arxiv.org/abs/2401.06449}{{\ttfamily arXiv:2401.06449
  [astro-ph.SR]}}.

\bibitem{meyer1989tool}
N.~Meyer-Vernet and C.~Perche, ``Tool kit for antennae and thermal noise near
  the plasma frequency,'' {\em Journal of Geophysical Research: Space Physics}
  {\bfseries 94} no.~A3, (1989) 2405--2415.

\bibitem{An:2024kls}
H.~An, S.~Ge, J.~Liu, and Z.~Lu, ``{Direct Detection of Dark Photon Dark Matter
  with the James Webb Space Telescope},''
  \href{http://arxiv.org/abs/2402.17140}{{\ttfamily arXiv:2402.17140
  [hep-ph]}}.

\bibitem{1989ApJ...337.1023C}
W.~A. {Coles} and J.~K. {Harmon}, ``{Propagation Observations of the Solar Wind
  near the Sun},'' \href{http://dx.doi.org/10.1086/167173}{{\em \apj}
  {\bfseries 337} (Feb., 1989) 1023}.

\bibitem{2001SSRv...97....9W}
R.~{Wohlmuth}, D.~{Plettemeier}, P.~{Edenhofer}, M.~K. {Bird}, A.~I. {Efimov},
  V.~E. {Andreev}, L.~N. {Samoznaev}, and I.~V. {Chashei}, ``{Radio Frequency
  Fluctuation Spectra During the Solar Conjunctions of the Ulysses and Galileo
  Spacecraft},'' \href{http://dx.doi.org/10.1023/A:1011845221808}{{\em Space
  Science Reviews} {\bfseries 97} (May, 2001) 9--12}.

\bibitem{2007ApJ...671..894T}
G.~{Thejappa}, R.~J. {MacDowall}, and M.~L. {Kaiser}, ``{Monte Carlo Simulation
  of Directivity of Interplanetary Radio Bursts},''
  \href{http://dx.doi.org/10.1086/522664}{{\em \apj} {\bfseries 671} no.~1,
  (Dec., 2007) 894--906}.

\bibitem{2008ApJ...676.1338T}
G.~{Thejappa} and R.~J. {MacDowall}, ``{Effects of Scattering on Radio Emission
  from the Quiet Sun at Low Frequencies},''
  \href{http://dx.doi.org/10.1086/528835}{{\em \apj} {\bfseries 676} no.~2,
  (Apr., 2008) 1338--1345}.

\bibitem{2018ApJ...857...82K}
V.~{Krupar}, M.~{Maksimovic}, E.~P. {Kontar}, A.~{Zaslavsky}, O.~{Santolik},
  J.~{Soucek}, O.~{Kruparova}, J.~P. {Eastwood}, and A.~{Szabo},
  ``{Interplanetary Type III Bursts and Electron Density Fluctuations in the
  Solar Wind},'' \href{http://dx.doi.org/10.3847/1538-4357/aab60f}{{\em \apj}
  {\bfseries 857} no.~2, (Apr., 2018) 82}.

\bibitem{2020ApJS..246...57K}
V.~{Krupar}, A.~{Szabo}, M.~{Maksimovic}, O.~{Kruparova}, E.~P. {Kontar}, L.~A.
  {Balmaceda}, X.~{Bonnin}, S.~D. {Bale}, M.~{Pulupa}, D.~M. {Malaspina}, J.~W.
  {Bonnell}, P.~R. {Harvey}, K.~{Goetz}, T.~{Dudok de Wit}, R.~J. {MacDowall},
  J.~C. {Kasper}, A.~W. {Case}, K.~E. {Korreck}, D.~E. {Larson}, R.~{Livi},
  M.~L. {Stevens}, P.~L. {Whittlesey}, and A.~M. {Hegedus}, ``{Density
  Fluctuations in the Solar Wind Based on Type III Radio Bursts Observed by
  Parker Solar Probe},'' \href{http://dx.doi.org/10.3847/1538-4365/ab65bd}{{\em
  The Astrophysical Journal Supplement Series} {\bfseries 246} no.~2, (Feb.,
  2020) 57}, \href{http://arxiv.org/abs/2001.03476}{{\ttfamily arXiv:2001.03476
  [astro-ph.SR]}}.

\bibitem{Brahma:2023zcw}
N.~Brahma, A.~Berlin, and K.~Schutz, ``{Photon-dark photon conversion with
  multiple level crossings},''
  \href{http://dx.doi.org/10.1103/PhysRevD.108.095045}{{\em Phys. Rev. D}
  {\bfseries 108} no.~9, (2023) 095045},
  \href{http://arxiv.org/abs/2308.08586}{{\ttfamily arXiv:2308.08586
  [hep-ph]}}.

\bibitem{1998SoPh..183..165L}
Y.~{Leblanc}, G.~A. {Dulk}, and J.-L. {Bougeret}, ``{Tracing the Electron
  Density from the Corona to 1 au},''
  \href{http://dx.doi.org/10.1023/A:1005049730506}{{\em Solar Physics}
  {\bfseries 183} no.~1, (Nov., 1998) 165--180}.

\bibitem{Pulupa2017TheProcessing}
M.~Pulupa, S.~D. Bale, J.~W. Bonnell, T.~A. Bowen, N.~Carruth, K.~Goetz,
  D.~Gordon, P.~R. Harvey, M.~Maksimovic, J.~C. Mart{\'{i}}nez-Oliveros,
  M.~Moncuquet, P.~Saint-Hilaire, D.~Seitz, and D.~Sundkvist, ``{The solar
  probe plus radio frequency spectrometer: Measurement requirements, analog
  design, and digital signal processing},''
  \href{http://dx.doi.org/10.1002/2016JA023345}{{\em Journal of Geophysical
  Research: Space Physics} {\bfseries 122} no.~3, (3, 2017) 2836--2854}.
  \url{https://onlinelibrary.wiley.com/doi/10.1002/2016JA023345}.

\bibitem{PSPwebPosition}
 \url{https://psp-gateway.jhuapl.edu/website/Tools/PositionCalculator/}.

\bibitem{2020Moncuquet}
M.~{Moncuquet}, N.~{Meyer-Vernet}, K.~{Issautier}, M.~{Pulupa}, J.~W.
  {Bonnell}, S.~D. {Bale}, T.~{Dudok de Wit}, K.~{Goetz}, L.~{Griton}, P.~R.
  {Harvey}, R.~J. {MacDowall}, M.~{Maksimovic}, and D.~M. {Malaspina}, ``{First
  In Situ Measurements of Electron Density and Temperature from Quasi-thermal
  Noise Spectroscopy with Parker Solar Probe/FIELDS},''
  \href{http://dx.doi.org/10.3847/1538-4365/ab5a84}{{\em The Astrophysical
  Journal Supplement Series} {\bfseries 246} no.~2, (Feb., 2020) 44},
  \href{http://arxiv.org/abs/1912.02518}{{\ttfamily arXiv:1912.02518
  [astro-ph.SR]}}.

\bibitem{PSP_QTN_CNES}
 \url{https://cdpp-archive.cnes.fr/}.

\bibitem{An:2022hhb}
H.~An, S.~Ge, W.-Q. Guo, X.~Huang, J.~Liu, and Z.~Lu, ``{Direct Detection of
  Dark Photon Dark Matter Using Radio Telescopes},''
  \href{http://dx.doi.org/10.1103/PhysRevLett.130.181001}{{\em Phys. Rev.
  Lett.} {\bfseries 130} no.~18, (2023) 181001},
  \href{http://arxiv.org/abs/2207.05767}{{\ttfamily arXiv:2207.05767
  [hep-ph]}}.

\bibitem{Cowan2011}
G.~Cowan, K.~Cranmer, E.~Gross, and O.~Vitells, ``{Asymptotic formulae for
  likelihood-based tests of new physics},''
  \href{http://dx.doi.org/10.1140/epjc/s10052-011-1554-0}{{\em The European
  Physical Journal C} {\bfseries 71} no.~2, (2, 2011) 1554}.
  \url{https://link.springer.com/article/10.1140/epjc/s10052-011-1554-0
  http://link.springer.com/10.1140/epjc/s10052-011-1554-0}.

\bibitem{ParticleDataGroup:2018ovx}
{\bfseries Particle Data Group} Collaboration, M.~Tanabashi {\em et~al.},
  ``{Review of Particle Physics},''
  \href{http://dx.doi.org/10.1103/PhysRevD.98.030001}{{\em Phys. Rev. D}
  {\bfseries 98} no.~3, (2018) 030001}.

\bibitem{zaslavsky2011antenna}
A.~Zaslavsky, N.~Meyer-Vernet, S.~Hoang, M.~Maksimovic, and S.~D. Bale, ``{On
  the antenna calibration of space radio instruments using the galactic
  background: General formulas and application to STEREO/WAVES},'' {\em Radio
  Science} {\bfseries 46} no.~2, (2011) .

\bibitem{An:2023mvf}
H.~An, S.~Ge, and J.~Liu, ``{Solar Radio Emissions and Ultralight Dark
  Matter},'' \href{http://dx.doi.org/10.3390/universe9030142}{{\em Universe}
  {\bfseries 9} no.~3, (2023) 142},
  \href{http://arxiv.org/abs/2304.01056}{{\ttfamily arXiv:2304.01056
  [hep-ph]}}.

\bibitem{2020Sci...369..694Y}
Z.~{Yang}, C.~{Bethge}, H.~{Tian}, S.~{Tomczyk}, R.~{Morton}, G.~{Del Zanna},
  S.~W. {McIntosh}, B.~B. {Karak}, S.~{Gibson}, T.~{Samanta}, J.~{He},
  Y.~{Chen}, and L.~{Wang}, ``{Global maps of the magnetic field in the solar
  corona},'' \href{http://dx.doi.org/10.1126/science.abb4462}{{\em Science}
  {\bfseries 369} no.~6504, (Aug., 2020) 694--697},
  \href{http://arxiv.org/abs/2008.03136}{{\ttfamily arXiv:2008.03136
  [astro-ph.SR]}}.

\bibitem{AxionLimits}
C.~O'Hare, ``cajohare/axionlimits: Axionlimits.''
  \url{https://cajohare.github.io/AxionLimits/}, July, 2020.

\bibitem{Bale2016TheTransients}
S.~D. Bale, K.~Goetz, P.~R. Harvey, P.~Turin, J.~W. Bonnell, T.~Dudok~de Wit,
  R.~E. Ergun, R.~J. MacDowall, M.~Pulupa, M.~Andre, M.~Bolton, J.~L. Bougeret,
  T.~A. Bowen, D.~Burgess, C.~A. Cattell, B.~D. Chandran, C.~C. Chaston, C.~H.
  Chen, M.~K. Choi, J.~E. Connerney, S.~Cranmer, M.~Diaz-Aguado, W.~Donakowski,
  J.~F. Drake, W.~M. Farrell, P.~Fergeau, J.~Fermin, J.~Fischer, N.~Fox,
  D.~Glaser, M.~Goldstein, D.~Gordon, E.~Hanson, S.~E. Harris, L.~M. Hayes,
  J.~J. Hinze, J.~V. Hollweg, T.~S. Horbury, R.~A. Howard, V.~Hoxie, G.~Jannet,
  M.~Karlsson, J.~C. Kasper, P.~J. Kellogg, M.~Kien, J.~A. Klimchuk, V.~V.
  Krasnoselskikh, S.~Krucker, J.~J. Lynch, M.~Maksimovic, D.~M. Malaspina,
  S.~Marker, P.~Martin, J.~Martinez-Oliveros, J.~McCauley, D.~J. McComas,
  T.~McDonald, N.~Meyer-Vernet, M.~Moncuquet, S.~J. Monson, F.~S. Mozer, S.~D.
  Murphy, J.~Odom, R.~Oliverson, J.~Olson, E.~N. Parker, D.~Pankow, T.~Phan,
  E.~Quataert, T.~Quinn, S.~W. Ruplin, C.~Salem, D.~Seitz, D.~A. Sheppard,
  A.~Siy, K.~Stevens, D.~Summers, A.~Szabo, M.~Timofeeva, A.~Vaivads, M.~Velli,
  A.~Yehle, D.~Werthimer, and J.~R. Wygant, ``{The FIELDS Instrument Suite for
  Solar Probe Plus: Measuring the Coronal Plasma and Magnetic Field, Plasma
  Waves and Turbulence, and Radio Signatures of Solar Transients},'' 12, 2016.
\newblock \url{https://link.springer.com/article/10.1007/s11214-016-0244-5}.

\end{thebibliography}\endgroup

\end{document}